\documentclass[12pt]{article}

\usepackage[reqno]{amsmath}
\usepackage{mathrsfs}
\usepackage{amssymb}

\usepackage{bbm}
\usepackage{epsfig}
\usepackage{array}
\usepackage{float}
\usepackage{color}
\usepackage{graphicx}

\usepackage{a4}
\usepackage{a4wide}
\usepackage{wasysym}

\def\ra{\rightarrow}
\def\gs{\mathrel{
   \rlap{\raise 0.511ex \hbox{$>$}}{\lower 0.511ex \hbox{$\sim$}}}}
\def\ls{\mathrel{
   \rlap{\raise 0.511ex \hbox{$<$}}{\lower 0.511ex \hbox{$\sim$}}}}
\newcommand{\obb}{0\mbox{$\nu\beta\beta$}}
\newcommand{\onbb}{neutrino-less double beta decay}
\newcommand{\ba}{\begin{array}{c}}
\newcommand{\baz}{\begin{array}{cc}}
\newcommand{\bad}{\begin{array}{ccc}}
\newcommand{\bav}{\begin{array}{cccc}}
\newcommand{\bea}{\begin{equation} \begin{array}{c}}
\newcommand{\eea}{ \end{array} \end{equation}}
\newcommand{\ea}{\end{array}}

\newcommand{\dms}{\mbox{$\Delta m^2_{\odot}$}}
\newcommand{\dma}{\mbox{$\Delta m^2_{\rm A}$}}
\newcommand{\meff}{\mbox{$\left| m_{ee} \right|$}}
\newcommand{\eV}{\mbox{ eV}}

\newcommand{\be}{\begin{eqnarray}}
\newcommand{\ee}{\end{eqnarray}}

\newcommand{\sss}{\sin^2 \theta_{12}}
\newcommand{\sch}{\sin^2 \theta_{13}}

\begin{document}

\begin{titlepage}
\title{\vspace*{-2.0cm} \hfill {\small TUM-HEP--623/06}\\[-5mm]
\hfill {\small hep--ph/0603111}\\[20mm]
\bf\Large
The Elements of the Neutrino Mass Matrix: Allowed Ranges and Implications 
of Texture Zeros 
\\[5mm]\ }

\author{
Alexander Merle\thanks{email: \tt alexander$\_$merle@ph.tum.de}~~,~~
Werner Rodejohann\thanks{email: \tt werner$\_$rodejohann@ph.tum.de} 
\\ \\
{\normalsize \it Physik-Department, Technische Universit\"at M\"unchen,}\\
{\normalsize \it  James-Franck-Strasse, D-85748 Garching, Germany}
}
\date{}
\maketitle
\thispagestyle{empty}

\begin{abstract}
\noindent
We study the range of the elements of the neutrino mass matrix $m_\nu$ 
in the charged lepton basis. Neutrino-less double beta decay is sensitive 
to the $ee$ element (the effective mass) of $m_\nu$.  
We then analyze the phenomenological implications of single texture zeros. 
In particular, interesting predictions for the effective mass 
can be obtained, in the sense that typically only little cancellation due to 
the Majorana phases is expected. 
Some cases imply constraints on the atmospheric neutrino mixing angle. 
\end{abstract}

\end{titlepage}

\section{\label{sec:intro}Introduction}

Neutrino physics aims to identify the form and origin 
of the neutrino mass matrix \cite{APSgen}. 
Towards this goal, literally dozens of new experiments are running, 
under construction or in development. 
For Majorana neutrinos, the neutrino mass matrix is given by 
\be
(m_\nu)_{\alpha \beta} = 
\left( U \, m_\nu^{\rm diag}  \, U^T \right)_{\alpha \beta}
\mbox{ with } \alpha, \beta = e, \mu, \tau~, 
\ee
where $m_\nu^{\rm diag}$ is a diagonal matrix containing the three 
mass states $m_{1,2,3}$. 
In the basis in which the 
charged lepton mass matrix is real and diagonal, 
$U$ is the leptonic mixing or PMNS matrix \cite{PMNS}. 
Being a symmetric matrix, $m_\nu$ contains  
six independent and complex entries, corresponding to nine physical 
parameters. 
Several analyzes \cite{ich,FS,barger} have been performed in 
order to study the form of the neutrino mass matrix as allowed 
by current data.  
Reconstructing $m_\nu$ is possible since for Majorana neutrinos 
-- in contrast to the quark sector -- the mass matrix is in general 
symmetric. 
In addition, the Majorana nature of neutrinos allows 
to probe an element of the mass matrix directly: namely, the 
second order weak interaction process 
$ (A,Z) \rightarrow (A,Z + 2) + 2 \, e^-$,  
denoted \onbb~(\obb), 
possesses a decay width proportional to the square of the so-called 
effective mass
$ \meff \equiv \left| \sum_i m_i \, U_{ei}^2 \right|$. 
In the charged lepton basis and if only the three light Majorana 
neutrinos as implied by the 
oscillation experiments are exchanged in \obb,  
this coherent sum is nothing but the $ee$ element of $m_\nu$. 
In order to experimentally reconstruct the mass matrix, 
the question arises if one can probe the other elements of $m_\nu$ in 
the same sense as the $ee$ element is probed by \obb. 
In principle, the answer is ``yes'': 
a mass matrix element 
$m_{\alpha \beta}$ will govern $\Delta L=2$ lepton number 
violating processes with the charged leptons $\alpha$ and $\beta$ 
in the final state. For instance, the rare decay  
$K^+ \ra \pi^+ \mu^- \mu^-$ has a branching ratio 
proportional to $|m_{\mu \mu}|^2$, where 
$m_{\mu \mu}$ is the $\mu\mu$ entry of $m_\nu$. 
However, the possibility to 
probe the mass matrix by these decays is only academic, since neutrino 
oscillation data predicts branching ratios of such processes 
up to 20 orders of magnitude below current experimental 
limits \cite{ich,kai,barger}. 

Another, often studied aspect of mass matrices is the possibility of 
texture zeros, which can serve as a tool to explain certain properties 
of the observed mass and mixing schemes\footnote{The structure of 
$m_\nu$ can also be constrained by assuming a vanishing determinant 
\cite{det,det1} or a vanishing trace \cite{trace}.}. 
Models generating such texture zeros can be based on 
the Froggatt--Nielsen mechanism \cite{FN} or certain 
flavor symmetries \cite{other}. 
Since texture zeros are very successful in the 
quark sector (see, e.g., \cite{quaaaaark} for a review), one expects such 
an approach to be useful for the lepton sector as well. 
It was found that in the charged lepton basis,  
and if neutrinos are Majorana particles,  
at most two zero entries in $m_\nu$ are allowed 
\cite{2zero,2zero1}\footnote{
For Dirac neutrinos there can be up to 5 zero entries \cite{dirac}.}. 
Interesting and observable correlations between the neutrino mass 
and mixing parameters follow from the presence of two zeros. 
Regarding the possibility of just one zero entry in $m_\nu$, 
we are only aware of analyzes making simplifying assumptions 
such as a vanishing determinant \cite{det1} 
or the equality of two entries \cite{hybrid}. 

In the present paper we wish to analyze the implications of vanishing 
mass matrix elements. 
Towards this goal, we study the individual mass matrix 
elements as functions of the smallest neutrino mass. Identifying the 
situations in which one of the entries can vanish, we obtain the 
phenomenological implications for the neutrino observables. 
This often concerns the $ee$ element, i.e., 
the predictions for \onbb~are modified 
by the constraint of one of the $m_{\alpha \beta}$ being zero. 
This modification concerns for instance the value of one of the 
Majorana phases, leading to little cancellation for the effective mass.\\  

We parameterize the PMNS matrix as 
\be 
\label{eq:Upara}
U = \left( \bad 
c_{12} c_{13} & s_{12} c_{13} & s_{13} \, e^{-i \delta}  \\[0.2cm] 
-s_{12} c_{23} - c_{12} s_{23} s_{13} e^{i \delta} 
& c_{12} c_{23} - s_{12} s_{23} s_{13} e^{i \delta} 
& s_{23} c_{13}  \\[0.2cm] 
s_{12} s_{23} - c_{12} c_{23} s_{13} e^{i \delta} & 
- c_{12} s_{23} - s_{12} c_{23} s_{13} e^{i \delta} 
& c_{23} c_{13}  \\ 
               \ea   \right) 
 {\rm diag}(1, e^{i \alpha}, e^{i (\beta + \delta)}) \, , 
\ee 
where we have used the usual notations $c_{ij} = \cos\theta_{ij}$, 
$s_{ij} = \sin\theta_{ij}$ and $\delta$ is the Dirac $CP$-violation 
phase, whereas $\alpha$ and $\beta$ are the two Majorana $CP$-violation 
phases \cite{BHP80}. 
In what regards the oscillation parameters, two independent 
mass squared differences, $\dms = m_2^2 - m_1^2$ and 
$\dma = |m_3^2 - m_1^2|$, as well as three mixing angles are 
currently under examination. 
Their best-fit, 1 and 3$\sigma$ values are \cite{valle} 
\begin{eqnarray}
\dms &=& 
\left(7.9^{+0.3\,, \,1.0}_{-0.3\,, \,0.8}\right) 
\cdot 10^{-5} \eV^2~,\nonumber\\
\sss &=& 0.31^{+0.02\,, \,0.09}_{-0.03\,, \,0.07} ~,\nonumber\\
\dma &=&  
\left(2.2^{+0.37\,, \,1.1}_{-0.27\,, \,0.8}\right) 
\cdot 10^{-3} \eV^2~,\\
\sin^2\theta_{23} &=& 0.50^{+0.06\,, \,0.18}_{-0.05\,, \,0.16} ~,\nonumber\\
\sch &<& 0.012~(0.046)~.\nonumber
\end{eqnarray}
The best-fit value for $\sch$ is 0. 

All six independent mass matrix elements 
depend crucially on the neutrino mass scheme, which is 
determined by the magnitude and the ordering (normal or inverted) of the 
individual neutrino masses: 
\be
\label{eq:masses}
\bav
\text{normal:}  & m_3 > m_2 > m_1 & \text{with} & m_2 = \sqrt{m_1^{2}+\dms} ~;
~~~~~~  m_3 = \sqrt{m_1^{2}+\dma} ~,\\[0.3cm]
\text{inverted:} & m_2 > m_1 > m_3 &  \text{with} & 
m_2 = \sqrt{m_3^{2}+\dms+\dma} ~;~~~~ 
m_1 = \sqrt{m_3^{2} + \dma} ~.
\ea
\ee
Of special interest are the following three extreme cases\footnote{In the 
last case, QD, we denote with $m_0$ the smallest neutrino mass, i.e., 
the definition written here applies for the normal ordering.}:  
\be \label{eq:nh}
\mbox{ normal hierarchy~(NH):} 
& ~~~~~~~~~~~~~ 
|m_3| \simeq \sqrt{\dma} \gg |m_{2}| \simeq \sqrt{\dms} \gg |m_1|~,\\[0.3cm]
\mbox{ inverted hierarchy~(IH):} \label{eq:ih}
& |m_2| \simeq |m_1| \simeq \sqrt{\dma} \gg |m_{3}| ~,\\[0.3cm]
\mbox{ quasi-degeneracy~(QD):} \label{eq:qd}
& ~~~~~~~~ m_0 \equiv |m_1| \simeq |m_2| \simeq |m_3|  \gg \sqrt{\dma} ~.
\label{eq:mass}
\ee
In the following Sections we will analyze the six independent 
mass matrix entries and study the implications of one of them 
being zero. 
It turns out that interesting implications from a texture zero in 
$m_{ee}$ only occur for NH (see Section \ref{sec:mee}). 
If $m_{e \mu}$ or $m_{e \tau}$ should 
vanish, then this yields interesting correlations in case of IH and QD 
(see Section \ref{sec:mem}). The case of the $\mu\mu$, $\tau\tau$ and 
$\mu\tau$ entries is studied in Section \ref{sec:mmm}. Only for 
quasi-degenerate neutrino masses there are notable correlations, which 
interestingly affect the atmospheric neutrino mixing angle $\theta_{23}$. 
We conclude and summarize in Section \ref{sec:concl}.\\

\section{\label{sec:mee}The Mass Matrix Element $m_{e e}$} 

The $ee$ entry of $m_\nu$ has been the subject of intense investigation in 
the past \cite{STP_rev} and we have nothing new to add here. 
For the sake of completeness, we nevertheless summarize 
the important features of $m_{ee}$. 
The current bound on \meff~is \cite{HM} 
\be \label{eq:current} 
\meff \le 0.35\, \zeta~{\rm eV}~, 
\ee
where we have indicated with the factor 
$\zeta={\cal O}(1)$ the uncertainty due to the calculation 
of the nuclear matrix elements of \obb. 
For an overview of the current status of \obb, see \cite{STP_rev,APSmass}. 
Given the fact that \meff{} depends on 7 out of 9 parameters 
contained in the neutrino mass matrix 
(it neither depends on $\delta$ nor on $\theta_{23}$), 
it is obvious that a huge 
amount of information would be encoded in an observation of \obb.

The effective mass \meff~reads in general and in the 
three extreme cases from Eqs.~(\ref{eq:nh}, \ref{eq:ih}, \ref{eq:qd}) 
as follows: 
\bea \label{eq:mee}
m_{e e} = c_{13}^2 
\left(m_1 \, c_{12}^2 +  e^{2 i \alpha} \, m_2  \, s_{12}^2 \right) + 
e^{2 i \beta} \, m_3  \, s_{13}^2 \\[0.3cm]
\Rightarrow \meff \simeq \left\{
\baz 
\left| e^{2i(\alpha - \beta)} \, \sqrt{\dms} \, c_{13}^2 \, s_{12}^2 + 
\sqrt{\dma} \, s_{13}^2 \right| 
& \mbox{ NH,} \\[0.2cm]
\sqrt{\dma} \, c_{13}^2 \, \sqrt{1 - \sin^2 2 \theta_{12} \, \sin^2 \alpha} 
& \mbox{ IH,} \\[0.2cm]
m_0 \, 
\left| 
c_{12}^2 + e^{2 i \alpha} \, s_{12}^2 + e^{2 i \beta} \, s_{13}^2 
\right| 
& \mbox{ QD.} 
\ea \right. 
\eea
A numerical evaluation of these expressions gives with best-fit, 
1$\sigma$ and 3$\sigma$ values of the oscillation parameters for 
NH 2.8 (1.8--3.5, 0--6.2) meV, for IH 0.02--0.05 (0.01--0.05, 0.01--0.06) eV 
and for QD with $m_0=0.5$ eV one gets 0.19--0.50 (0.16--0.51, 0.08--0.52) 
eV. 

\begin{figure}[tb]
\hspace{-1.2cm}
\begin{tabular}[h]{lr}\hspace{-1.2cm}
\epsfig{file=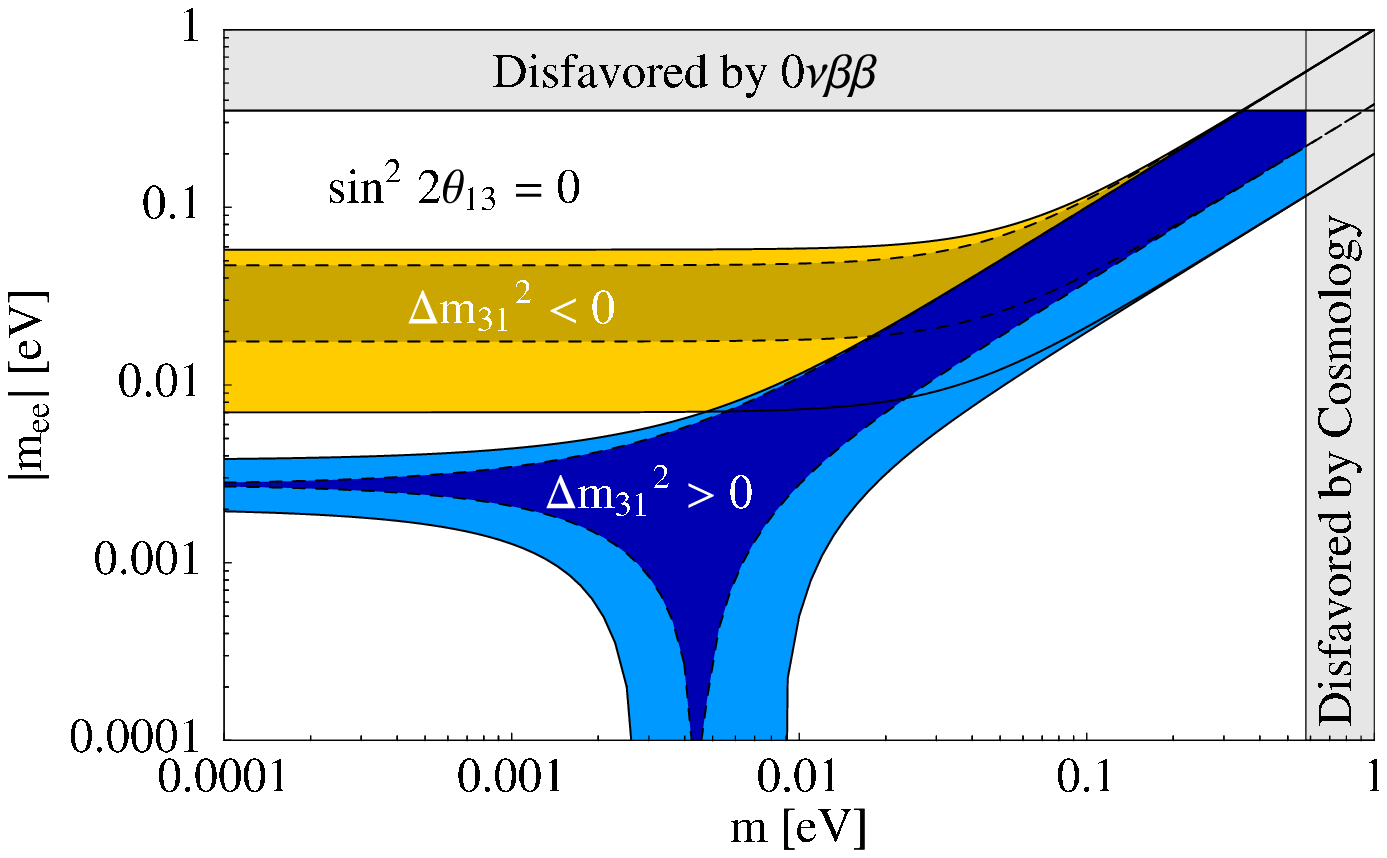,width=9.5cm,height=7cm} & \hspace{-1cm}
\epsfig{file=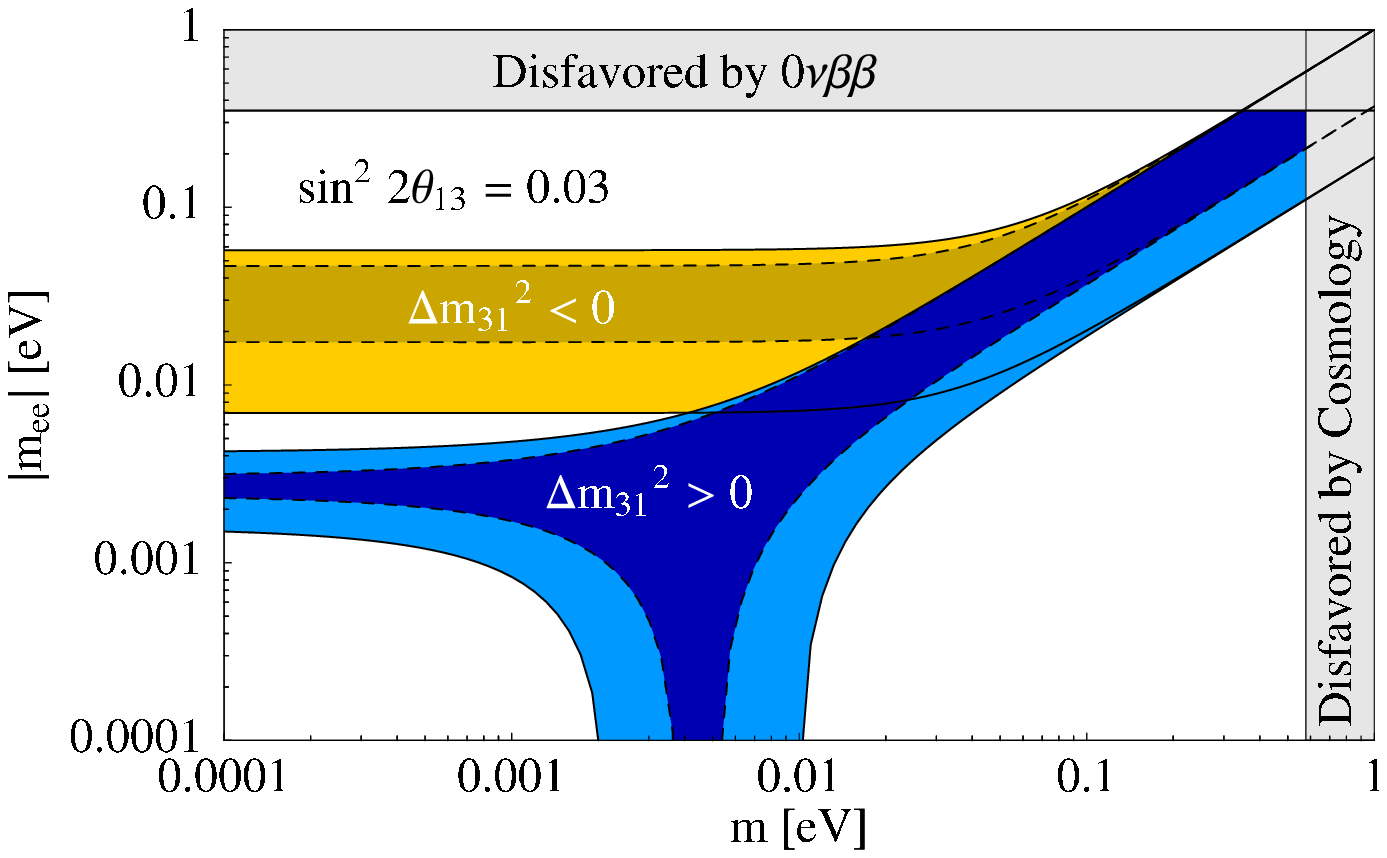,width=9.5cm,height=7cm} \\ \hspace{-1.2cm}
\epsfig{file=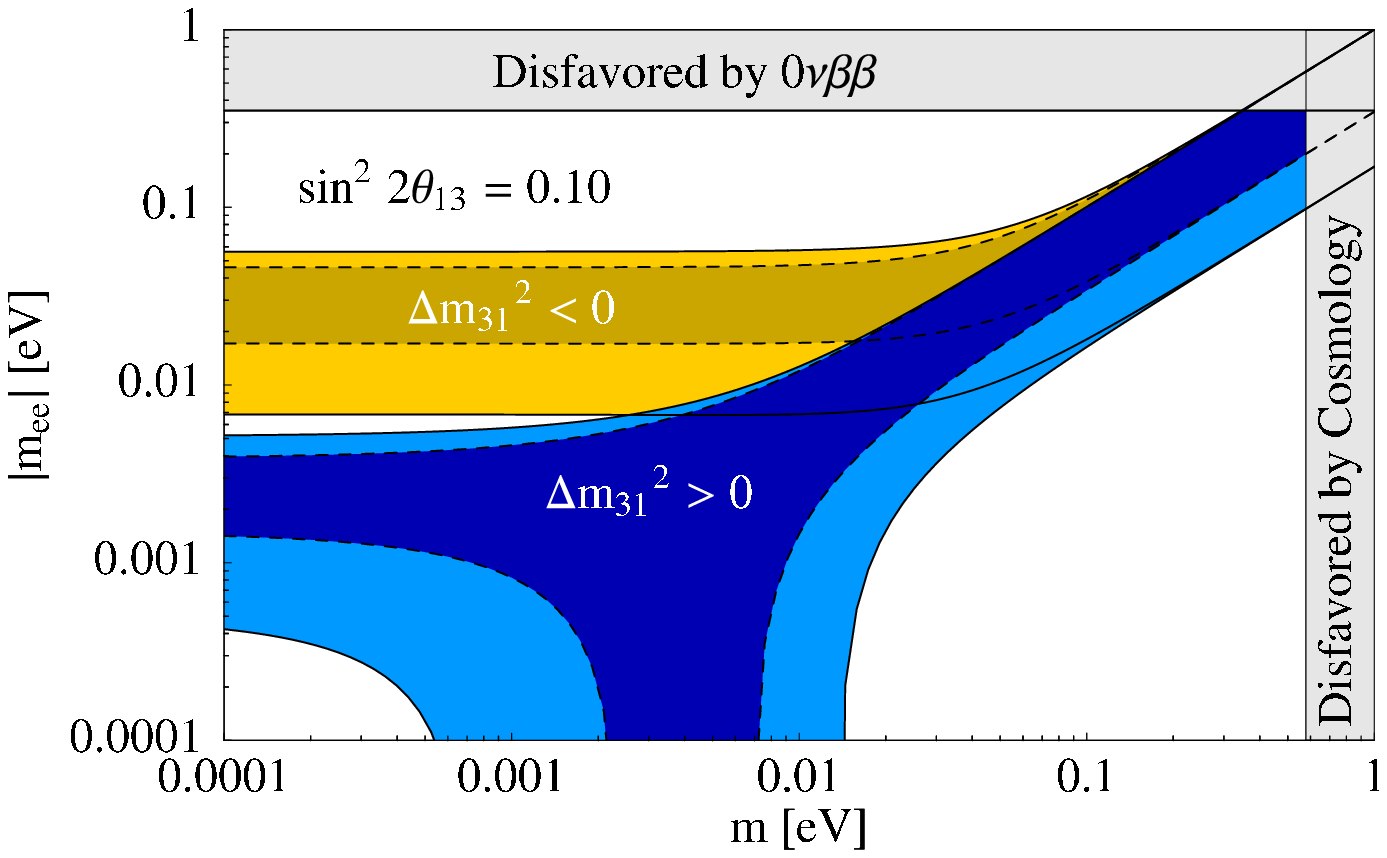,width=9.5cm,height=7cm} & \hspace{-1cm}
\epsfig{file=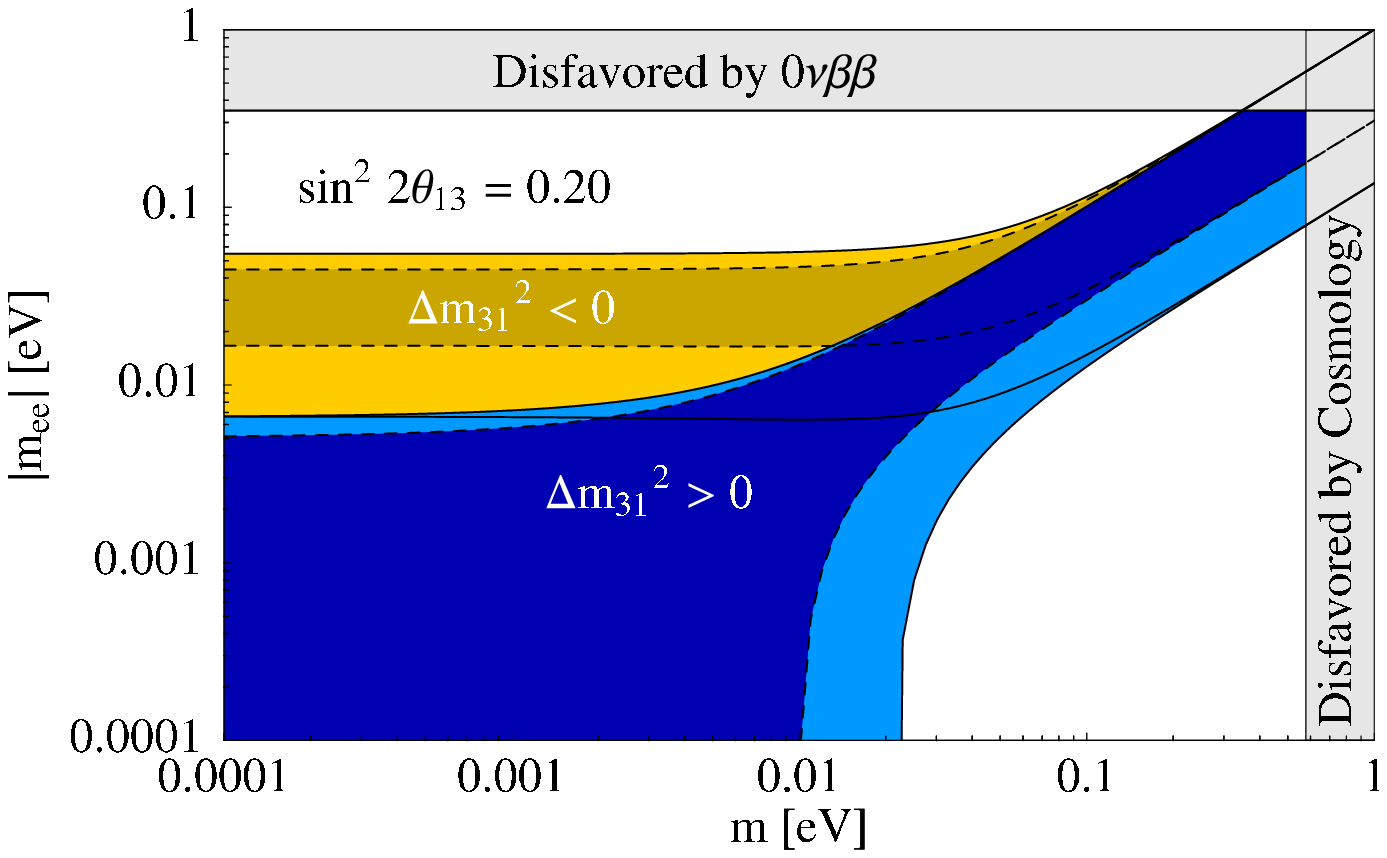,width=9.5cm,height=7cm}
\end{tabular}
\caption{\label{fig:mee}The absolute value of the mass matrix 
element $m_{ee}$ against the 
smallest neutrino mass for the normal and inverted mass ordering for 
four representative values of $\theta_{13}$. The 
best-fit and $3\sigma$ ranges of the oscillation parameters are used.}
\end{figure}

Interestingly, it will turn out that from all 
six independent entries in the mass matrix, the $ee$ element 
as given in Eq.~(\ref{eq:mee}) looks most simple. 
Obviously, this has its reason in the simple form of the $U_{ei}$ entries 
in the PMNS matrix, and is therefore a consequence of the chosen 
parametrization. 
For $\theta_{13} = 0$ it is given by 
\be \label{eq:mee_hie}
(m_{ee})_{\theta_{13}=0} 
= m_1 \, c_{12}^2 +  e^{2 i \alpha} \, m_2  \, s_{12}^2 
\Rightarrow \meff_{\theta_{13}=0}  \simeq \left\{ 
\baz 
\sqrt{\dms} \, s_{12}^2 
& \mbox{ NH,} \\[0.2cm]
\sqrt{\dma} \, \sqrt{1 - \sin^2 2 \theta_{12} \, \sin^2 \alpha} 
& \mbox{ IH,} \\[0.2cm]
m_0 \, \sqrt{1 - \sin^2 2 \theta_{12} \, \sin^2 \alpha} 
& \mbox{ QD.} 
\ea \right. 
\ee
The scale of $m_{ee}$ for zero $\theta_{13}$ and NH is 
roughly $\sqrt{\dms} \, s_{12}^2 \simeq 0.003 $ eV, for IH it is 
$\meff \simeq \sqrt{\dma} \simeq 0.04$ eV and for QD one has 
$\meff \simeq m_0$ (cancellations for IH and QD can be of order 50 $\%$, 
though). 
Plugging in the best-fit values, 1$\sigma$ and 3$\sigma$ ranges, 
$|m_{ee}|$ is given by  
2.8 (2.4--3.0, 2.0--3.8) meV for NH, 17.8--46.9 (14.9--50.7, 7.5--57.4) meV 
for IH and, for QD with 
$m_0=0.5$ eV, 0.19--0.50 (0.17--0.50, 0.10--0.50) eV. 
The Majorana phase $\alpha$ is crucial for the magnitude of \meff~in 
case of IH and QD. When $\sin \alpha = 0$, then there is no cancellation, 
whereas for $\sin \alpha = 1$ maximal cancellation occurs. 
Moreover, $\sin \alpha = 0$ implies sizable instability 
with respect to radiative corrections. 
In Fig.\ \ref{fig:mee} we show the $ee$ element as a function of the 
smallest neutrino mass for four characteristic values of $\theta_{13}$. 
In this and some of the following plots we indicate a typical 
cosmological limit on the sum of neutrino masses
$\Sigma \equiv \sum m_i $ of 1.74~eV (thus $m < 0.58$~eV 
for the lightest neutrino mass), obtained by an analysis of SDSS and 
WMAP data \cite{Tegmark}.\\

Turning to the possibility of texture zeros, it is 
well known that $m_{ee}=0$ is only possible for a limited parameter 
range in the normal mass ordering but for all values of 
$\theta_{13}$ \cite{mee0,LMR}. 
As emphasized recently in \cite{LMR}, and reproduced here in 
Fig.\ \ref{fig:mee}, current data implies that 
not too large $\theta_{13}$ values and $m_1$ between 
$10^{-3}$ and $10^{-2}$ eV leads to vanishing \meff~(the ``chimney''). For 
rather large $\sin^2 2\theta_{13} \gs 0.15$, all values 
of $m_1 \ls 0.01$ eV can lead to $\meff=0$. 
For $m_1=0$ and $\cos 2(\alpha - \beta) = -1$ 
the effective mass vanishes for rather large values of 
$\theta_{13}$, namely 
\be 
\sin^2 2 \theta_{13}  \simeq 4 \, 
\sin^2 \theta_{12} \, \sqrt{\frac{\dms}{\dma}}~,
\ee 
whose best-fit value is 0.24  
(1$\sigma$: 0.19--0.28; 3$\sigma$: 0.14--0.40). 
If $\theta_{13}=0$ and $\cos 2 \alpha = -1$, then 
\be
m_1 = \sin^{2} \theta_{12} \, \sqrt{\frac{\dms}{\cos 2 \theta_{12}}}
\ee 
leads to $\meff=0$.  The best-fit value of $m_1$ is 4.5~meV 
(1$\sigma$: 3.7--5.1~meV, 3$\sigma$: 2.8--8.4~meV). 
The implied values of the sum of neutrino masses are smaller than current 
limits from cosmology, but merely one order of magnitude below.  
Next generation observations \cite{steen} will probe such values.

\section{\label{sec:mem}The Mass Matrix Elements $m_{e \mu}$ and 
 $m_{e \tau}$}

\begin{figure}[tb]
\hspace{-1.2cm}
\begin{tabular}[h]{lr}\hspace{-1.2cm}
\epsfig{file=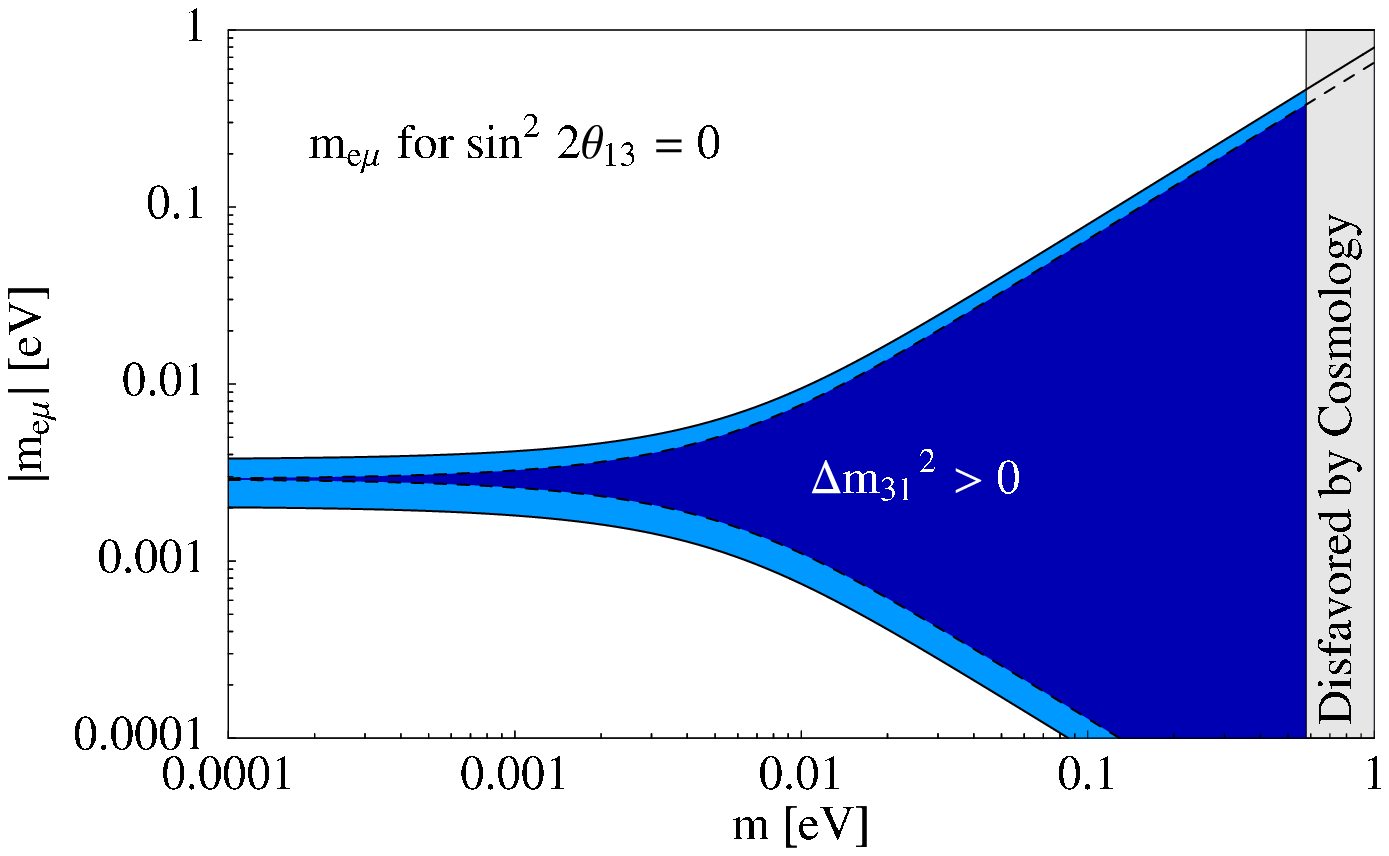,width=9.5cm,height=6.7cm}  & \hspace{-1cm} 
\epsfig{file=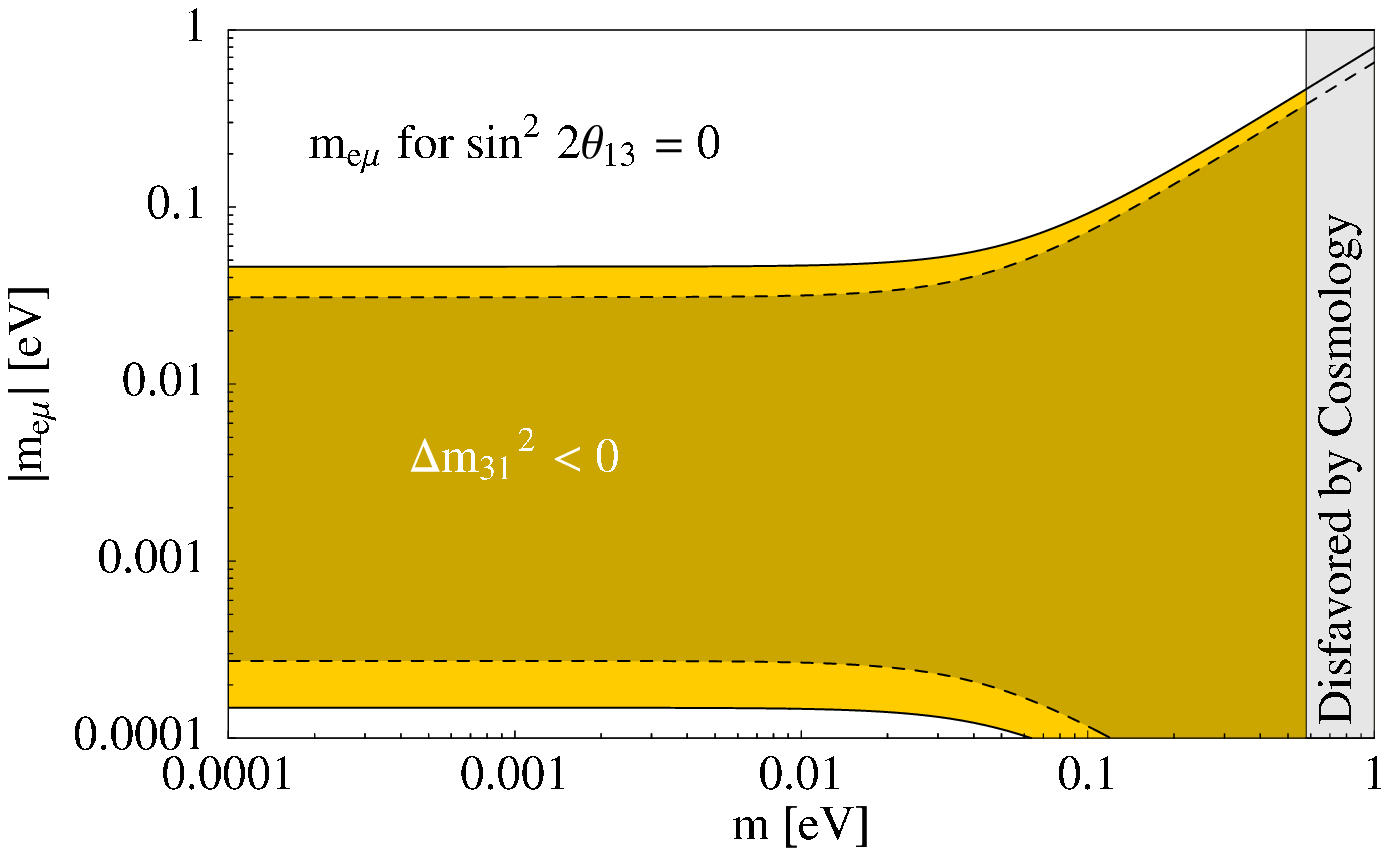,width=9.5cm,height=6.7cm} \\ \hspace{-1.2cm}
\epsfig{file=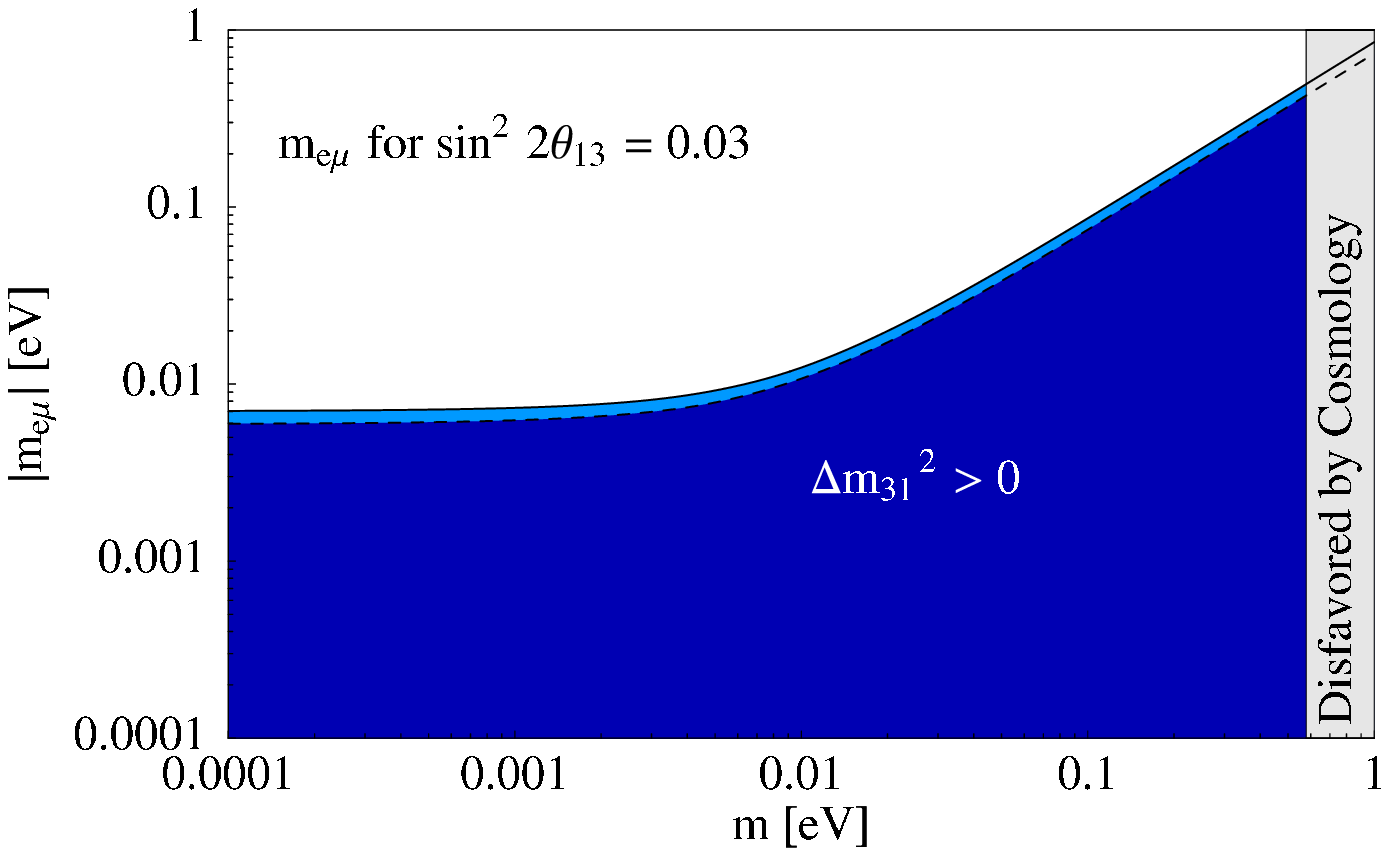,width=9.5cm,height=6.7cm} & \hspace{-1cm} 
\epsfig{file=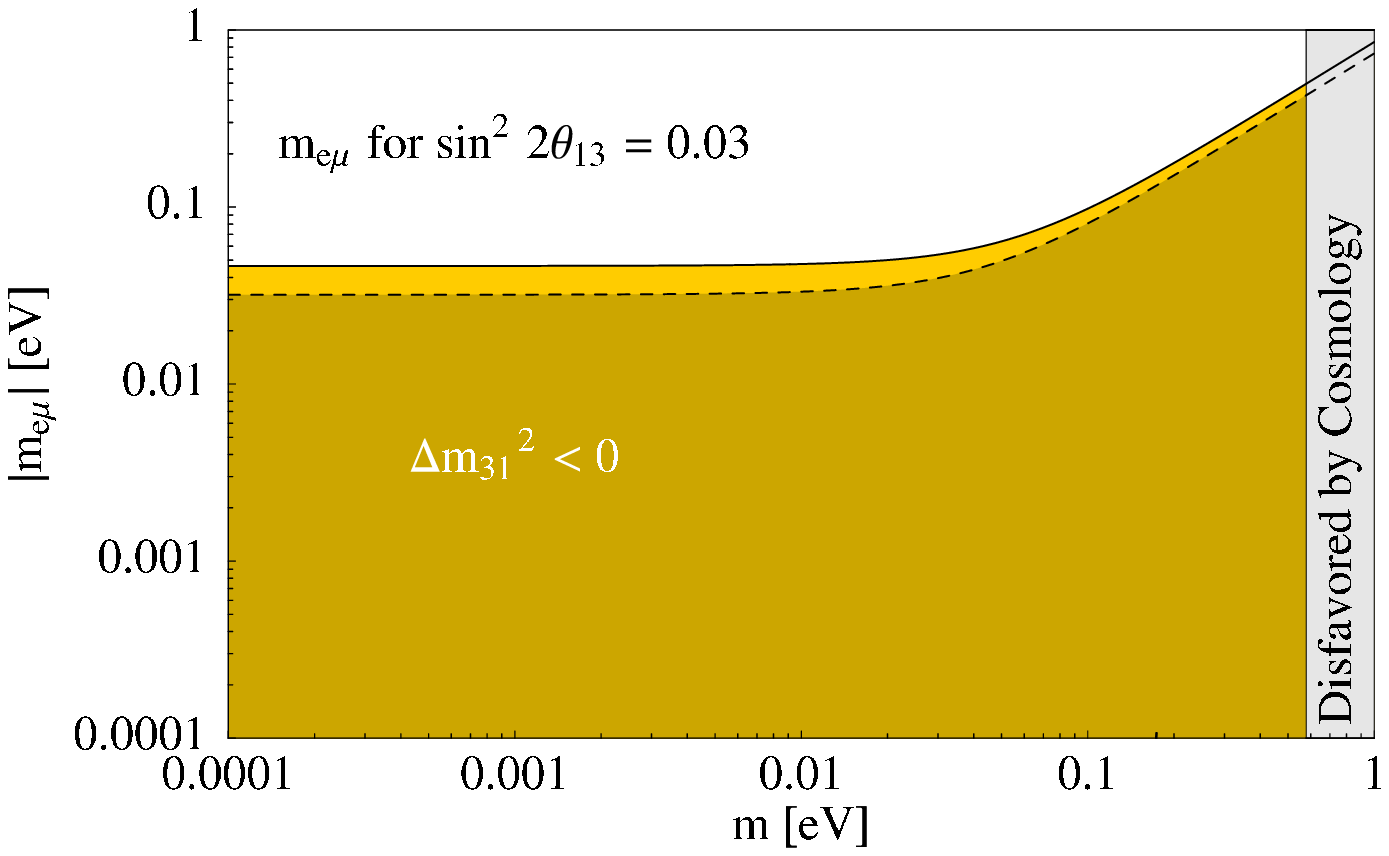,width=9.5cm,height=6.7cm} \\ \hspace{-1.2cm}
\epsfig{file=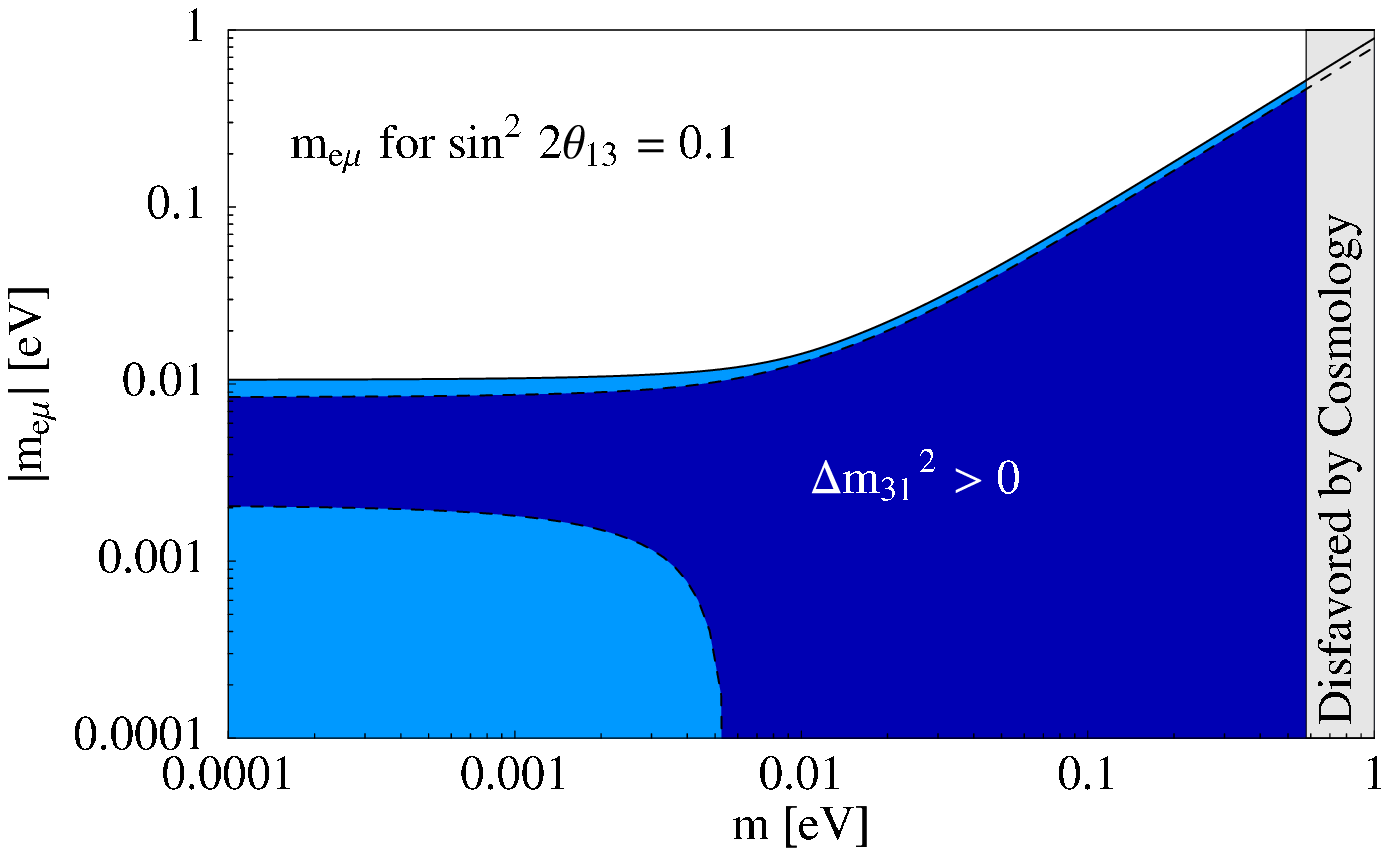,width=9.5cm,height=6.7cm} & \hspace{-1cm} 
\epsfig{file=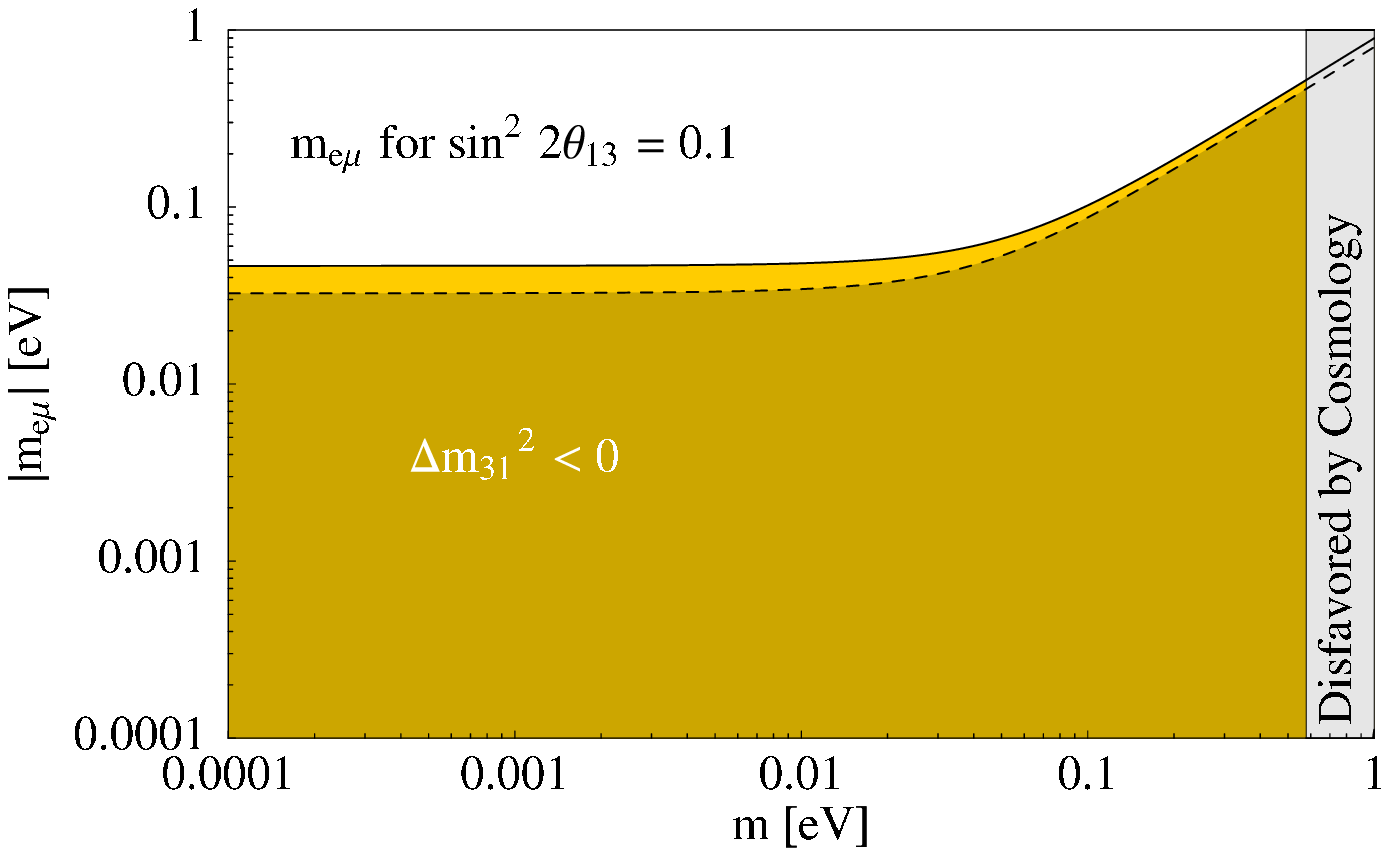,width=9.5cm,height=6.7cm}
\end{tabular}
\caption{\label{fig:mem}
The absolute value of the mass matrix 
element $m_{e\mu}$ against the 
smallest neutrino mass for the normal and inverted mass ordering for 
three representative values of $\theta_{13}$. The normal (inverted) mass 
ordering is given on the left (right) side. The 
best-fit and $3\sigma$ ranges of the oscillation parameters are used. 
The corresponding plot for $m_{e\tau}$ looks basically identical.}
\end{figure}

In general the $e\mu$ element of $m_\nu$ is given by 
\be \label{eq:mem}
m_{e \mu} = 
c_{13} \left( \left( e^{2i\alpha}\,m_2 - m_1 \right)
       s_{12}\,c_{12}\,c_{23} + e^{i \delta}
      \left( e^{2i \beta}\,m_3 - e^{2i  \alpha}\,m_2\,
         s_{12}^2 - 
        m_1\,c_{12}^2 \right) s_{23} \, s_{13} 
      \right)~.
\ee
Varying the three phases between $-\pi$ and $\pi$ and using the best-fit 
values and 3$\sigma$ ranges of the oscillation parameters 
as input we plot this element as a function of the smallest 
mass in Fig.\ \ref{fig:mem}.

\begin{figure}[tb]
\begin{center}
\epsfig{file=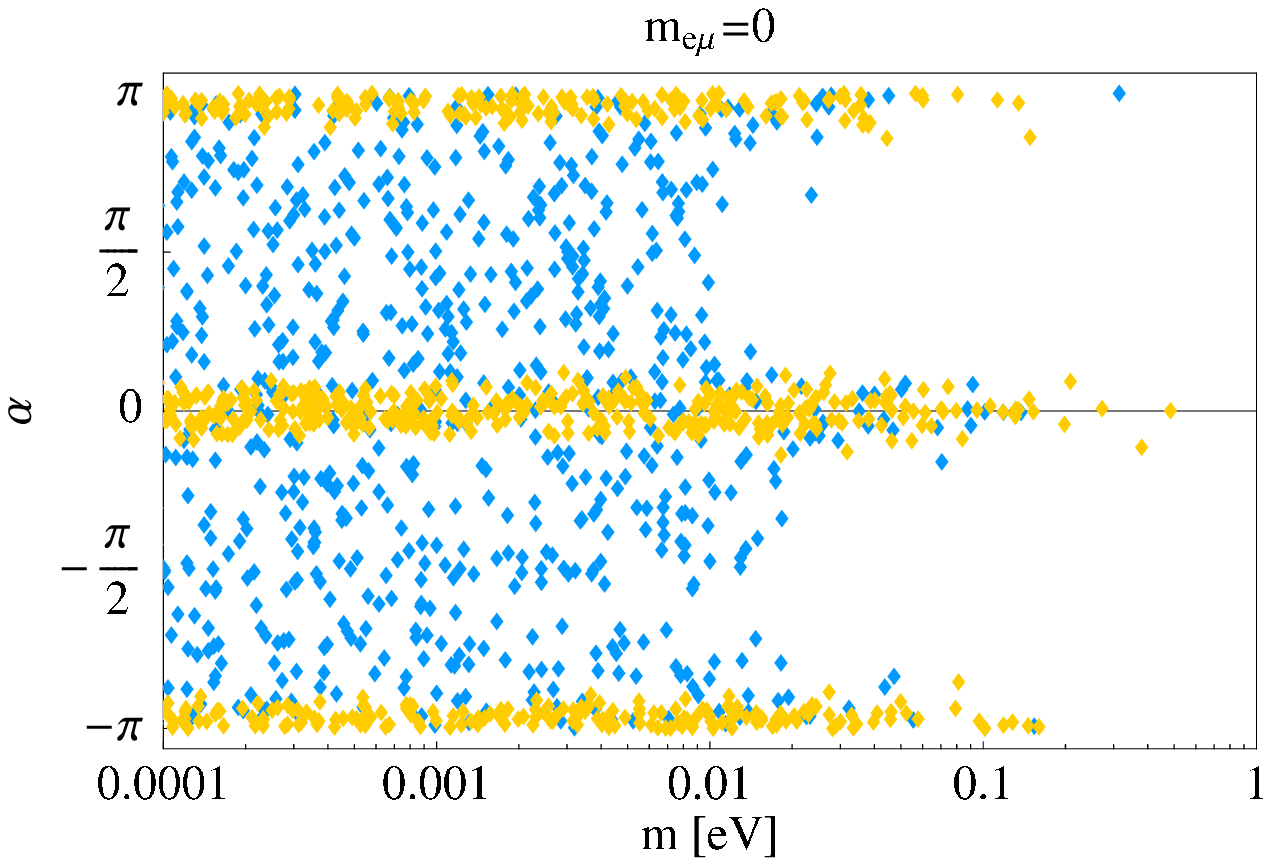,width=10cm}
\epsfig{file=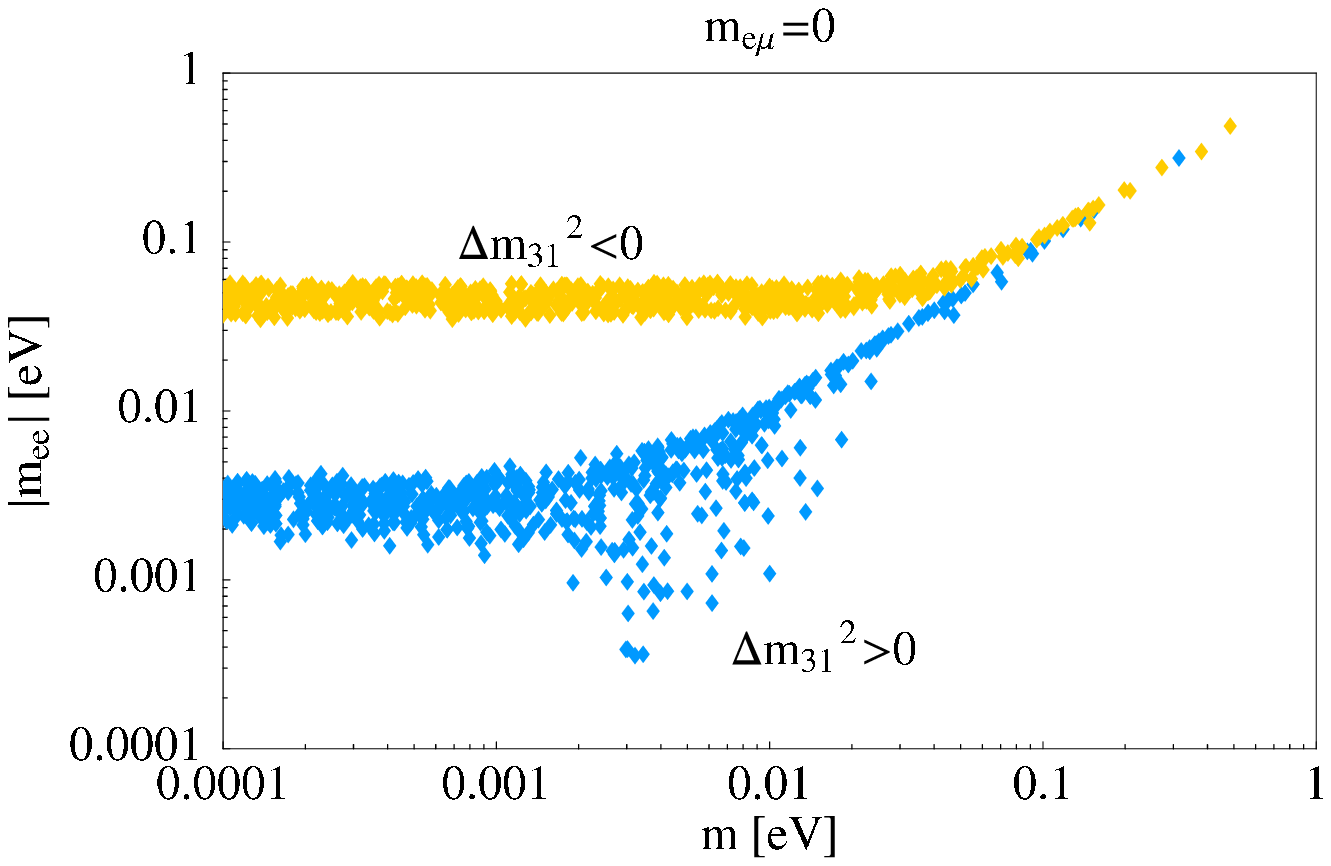,width=10cm}
\caption{\label{fig:corr_mem}Neutrino observables for 
$m_{e \mu}=0$. The upper plot is the smallest mass against the Majorana 
phase $\alpha$ and the lower plot is the smallest mass against 
the effective mass. The blue (dark) dots are for the normal ordering, 
the yellow (light) dots for the inverted ordering. 
We varied the oscillation parameters within their 3$\sigma$ 
ranges.}
\end{center}
\end{figure}

It is always helpful to consider certain extreme cases. In this and the 
following Sections, we will analyze the cases $\theta_{13}=0$ and 
$\theta_{23}=\pi/4$. 
\begin{itemize}
\item $\theta_{13}=0$: 
\be \label{eq:mem_t130}
(m_{e \mu})_{\theta_{13}=0} = \left( e^{2i\alpha}\,m_2 - m_1 \right) 
s_{12}\,c_{12}\,c_{23} 
\ee
This formula indicates that there is a lower limit 
on the $e\mu$ element:
\bea
|m_{e \mu}|_{\theta_{13}=0} \gs 
\left\{ 
\baz 
s_{12}\,c_{12}\,c_{23} \, \sqrt{\dms}  & \mbox{ NH,} \\[0.3cm]
s_{12}\,c_{12}\,c_{23} \, 
\frac{\Delta m^2_\odot}{ 2\sqrt{\Delta m^2_{\rm A}}} & \mbox{ IH,} \\[0.3cm]
s_{12}\,c_{12}\,c_{23} \, 
\frac{\Delta m^2_\odot}{2 m_0} & \mbox{ QD.}
\ea
\right. 
\eea
For the best-fit values this number is roughly 
0.003 eV for NH, 0.0003 eV for IH, and 
for QD with $m_0 = 0.5$ eV we have $2.6 \cdot 10^{-5}$ eV. 
There is also an upper limit for IH and QD, reading 
$2s_{12}\,c_{12}\,c_{23} \, \sqrt{\dma}$ and 
$2s_{12}\,c_{12}\,c_{23} \, m_0$, respectively. 
We conclude that for $\theta_{13}=0$ the $e\mu$ entry cannot vanish, 
but has a very low lower limit. 
Ignoring this small effect, we have 
\bea \label{eq:mem_t13}
|m_{e \mu}|_{\theta_{13}=0} \simeq \left\{ 
\baz 
\frac 12 \sqrt{\dms} \, c_{23} \, \sin 2 \theta_{12} 
& \mbox{ NH,} \\[0.2cm] 
c_{23} \, \sin 2 \theta_{12} \, 
\sqrt{\dma} \, \sin \alpha 
& \mbox{ IH,} \\[0.2cm] 
m_0 \, c_{23} \, \sin 2 \theta_{12} \, \sin \alpha
& \mbox{ QD.} 
\ea \right. 
\eea
The scale of $m_{e\mu}$ for zero $\theta_{13}$ and NH is 
roughly $\sqrt{\dms}/\sqrt{8}\simeq 0.002 $ eV, for IH and 
QD there can be almost total cancellation. 
Numerically, we have for the best-fit values and the 1 and 3$\sigma$ 
ranges 
2.9 (2.6--3.2, 2.0--3.8) meV for NH, 
0--30.7 (0--35.4, 0--45.7) meV for IH and for QD with $m_0 = 0.5$ eV one 
gets 0--0.33 (0--0.35, 0--0.40) eV.
The range for $m_{e \mu}$ in case of IH and QD is much larger than for NH, 
because -- in addition to the oscillation parameters -- 
the mass matrix element is proportional to $\sin \alpha$. 

\item $\theta_{23}=\pi/4$: 
\bea \label{eq:mem_t23}\hspace{-1cm}
(m_{e \mu})_{\theta_{23}=\pi/4} = \frac{c_{13}}{\sqrt{2}} 
\,\left( \left( e^{2i\alpha}\,m_2 - m_1 \right)
       s_{12}\,c_{12} + e^{i \delta}
      \left( e^{2i \beta}\,m_3 - e^{2i  \alpha}\,m_2\,
         s_{12}^2 - 
        m_1\,c_{12}^2 \right) s_{13}\,
       \right)~\\[0.3cm]\hspace{-0.61cm}
\Rightarrow |m_{e \mu}|_{\theta_{23}=\pi/4} \simeq \left\{ 
\baz 
\frac{c_{13}}{\sqrt{2}} 
\, \left| e^{2i\alpha}\,\sqrt{\dms} \, s_{12}\,c_{12} 
+ e^{i \delta} \, \left( e^{2i\beta} \sqrt{\dma} - 
e^{2i\alpha}\,\sqrt{\dms} \, s_{12}^2 \right) s_{13} 
\right| & \mbox{ NH,}\\[0.2cm]
\sqrt{\dma} \, 
\frac{c_{13}}{\sqrt{2}} \, \left| 
\left(e^{2 i \alpha} - 1\right) \, c_{12} \, s_{12} - e^{i \delta} 
\left(c_{12}^2 + e^{2 i \alpha} \, s_{12}^2 \right) s_{13} 
\right|  
& \mbox{ IH,}\\[0.2cm] 
m_0 \, 
\frac{c_{13}}{\sqrt{2}} \, \left| 
\left(e^{2 i \alpha} - 1\right) \, c_{12} \, s_{12} + e^{i \delta} 
\left(e^{2 i \beta} - c_{12}^2 - e^{2 i \alpha} \, s_{12}^2  \right) s_{13} 
\right| 
& \mbox{ QD.} 
\ea 
\right. 
\eea
Numerically, we have for the best-fit values and the 1 and 3$\sigma$ 
ranges 
2.9 (0--7.1, 0--12.3) meV for NH, 0--30.7 (0--34.8, 0--40.6) meV 
for IH and for QD with 
$m_0=0.5$ eV one gets 0--0.33 (0--0.38, 0--0.43) eV. 
In Fig.\ \ref{fig:mem} one can see that for NH and the best-fit 
oscillation parameters the value 
$\sin^2 2\theta_{13} = 0.03$ allows for complete cancellation for all 
allowed values of $m_1$. 
When $m_1$ is negligible, this can be understood by noting 
that when $\sin^2 2\theta_{13} = 0.03$, the two leading 
terms in the formula 
Eq.\ (\ref{eq:mem_t23}), namely $\sqrt{\dms} \, s_{12}\,c_{12}$ and 
$\sqrt{\dma} \, s_{13}$, happen to be almost identical. For 
smaller (larger) $\theta_{13}$ the first (second) term dominates and no 
cancellation for hierarchical neutrinos is possible. 
In case of IH and QD the $e\mu$ element can always vanish. This requires 
the first term in Eq.\ (\ref{eq:mem_t23}) to be small, thereby implying 
$\sin \alpha \simeq 0$, which leads to 
little cancellation in \obb~(see Eq.~(\ref{eq:mee})): 
\be \label{eq:meff_mem_t13}
|m_{ee}|_{\theta_{13}\neq0}^{m_{e\mu}=0} \simeq 
\left\{ 
\baz 
\sqrt{\dma} \, c_{13}^2  & \mbox{ IH,}\\[0.2cm]
m_0 \, c_{13}^2 & \mbox{ QD.} 
\ea
\right. 
\ee

\end{itemize}
We plot in Fig.~\ref{fig:corr_mem} the correlation between 
some of the observables which result if $m_{e\mu}=0$. All 
oscillation parameters are varied in their $3\sigma$ ranges. 
The upper plot is a scatter plot of the smallest mass against the Majorana 
phase $\alpha$ for both mass orderings.  As expected from the above 
considerations, for an inverted ordering and for 
quasi-degenerate masses $\sin \alpha$ is very small.  
Consequently, the effective mass is basically given by 
$\sqrt{\dma}$ or $m_0$, and the small band of values of \meff~for 
large masses is a consequence of 
$\sin \alpha \simeq 0$. Moreover, 
the gap between the maximal effective mass for NH 
and the minimal effective mass for IH is sizably larger than without 
the restriction $m_{e\mu}=0$ (cf.\ the lower plot of Fig.~\ref{fig:corr_mem}
with Fig.\ \ref{fig:mee}). This will make it easier to distinguish 
the normal from the inverted hierarchy via \obb.  
Note that $m_{ee}$ and $m_{e\mu}$ can vanish simultaneously 
\cite{2zero,2zero1} . 
The resulting correlation of parameters in this case 
and in all other allowed two zero cases is for the sake of 
completeness reproduced in Table \ref{tab:FGM}.\\

The results for the $e\tau$ element of $m_\nu$ are very similar to the 
ones for $m_{e\mu}$: the explicit form of $m_{e\tau}$ is 
\be \label{eq:met}
m_{e \tau} = 
c_{13} \left( \left( m_1 - e^{2i\alpha}\,
         m_2 \right) \,s_{12}\,c_{12} \,s_{23} + e^{i \delta}\,
      \left( e^{2i \beta}\,m_3 - e^{2i \alpha}\,m_2\,
         s_{12}^2 - m_1\,c_{12}^2  \right) c_{23} \, s_{13}  \right)~,
\ee
i.e., it is obtained from $m_{e \mu}$ by exchanging $s_{23}$ with $c_{23}$ 
and $c_{23}$ with $-s_{23}$. 
A plot of $m_{e\tau}$ against the smallest neutrino mass is basically 
indistinguishable from the plot of $m_{e\mu}$ against the smallest mass. 
This is the obvious consequence of the (approximate) 
$\mu$-$\tau$ exchange symmetry \cite{mutau} of the neutrino 
mass matrix in the charged lepton basis.

\section{\label{sec:mmm}The Mass Matrix Elements $m_{\mu\mu}$, 
$m_{\tau\tau}$ and $m_{\mu\tau}$}

\begin{figure}[tb]
\hspace{-1.2cm}
\begin{tabular}[h]{lr}\hspace{-1.2cm}
\epsfig{file=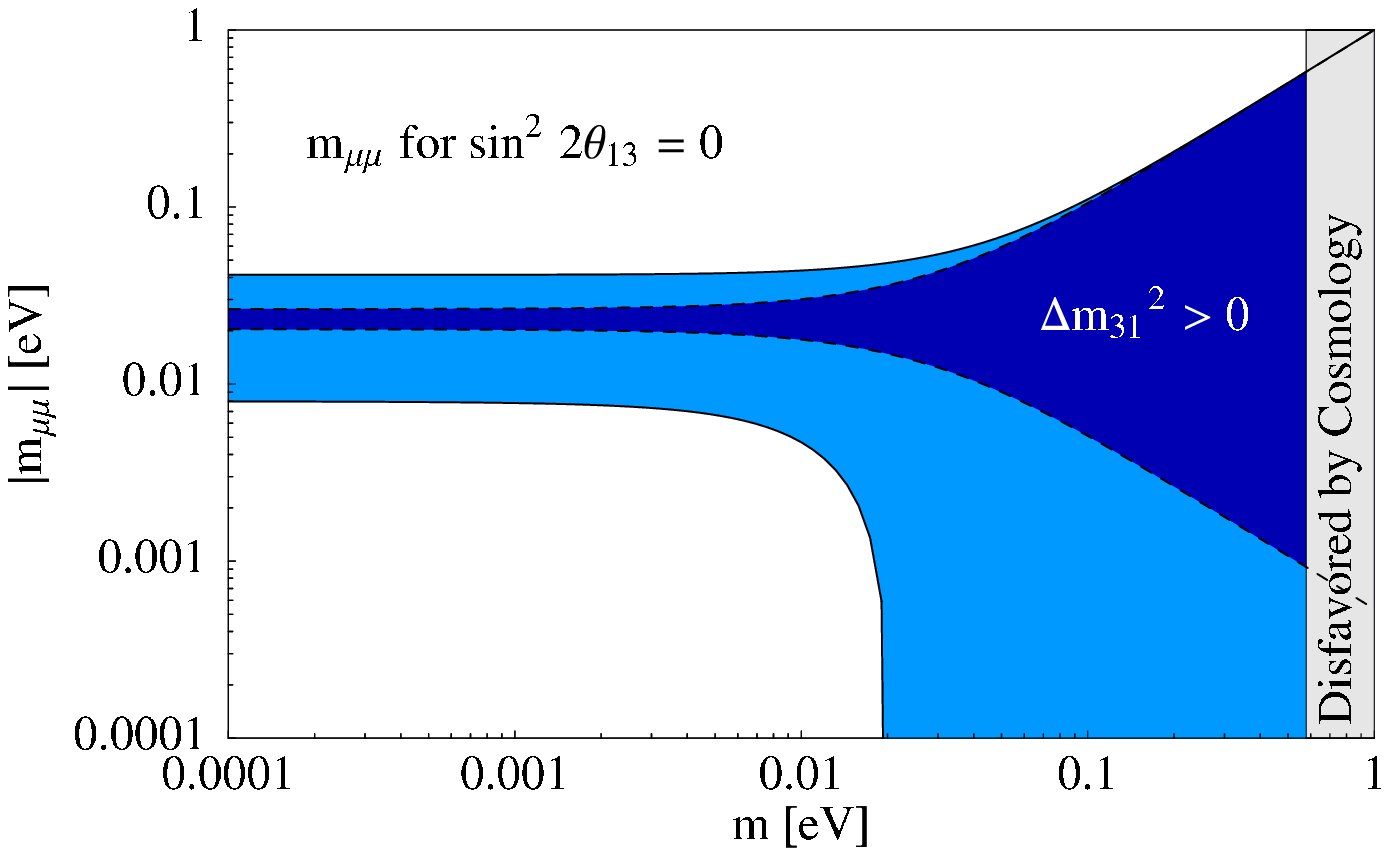,width=9.5cm,height=6.7cm} & \hspace{-1cm} 
\epsfig{file=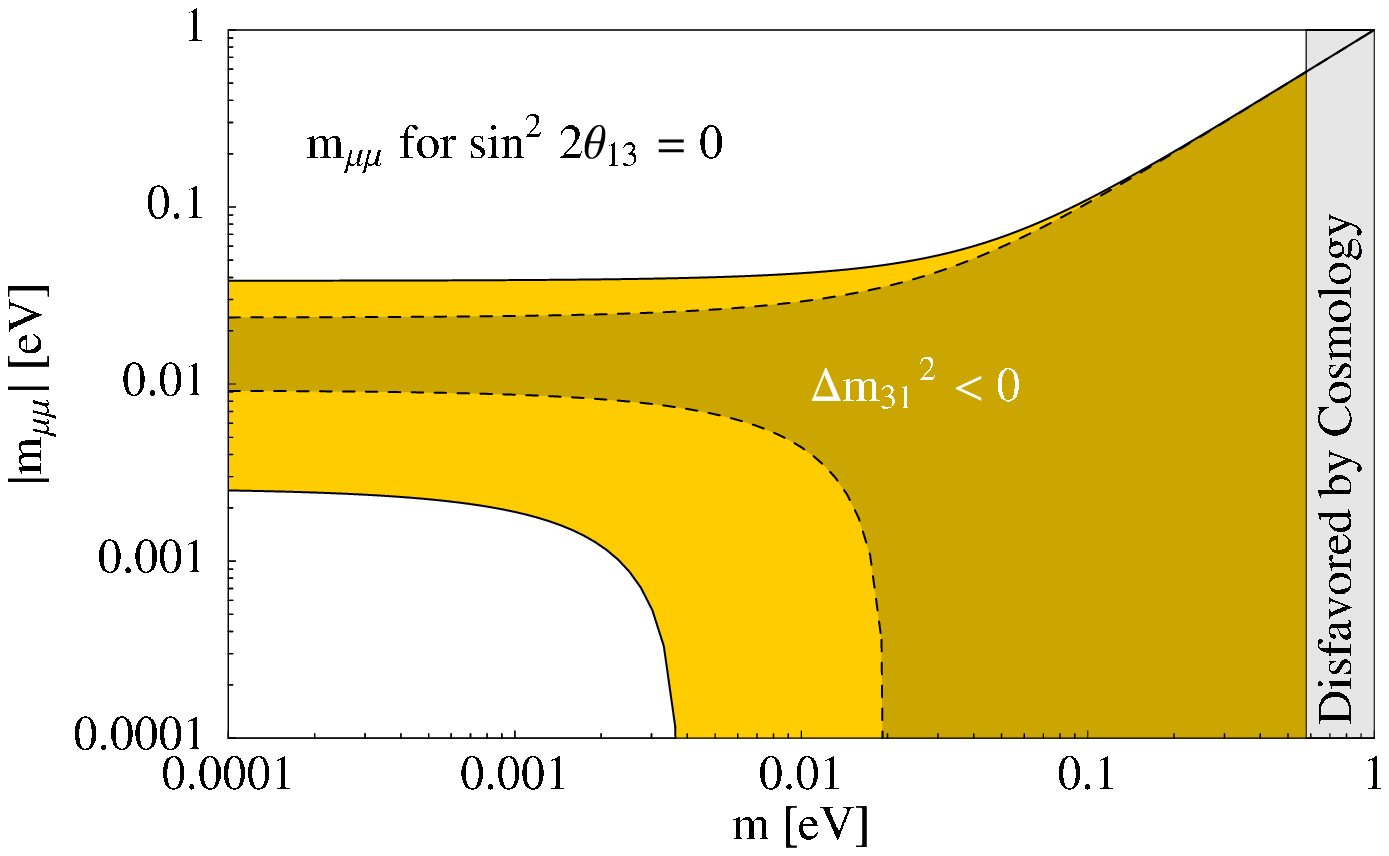,width=9.5cm,height=6.7cm} \\ \hspace{-1.2cm}
\epsfig{file=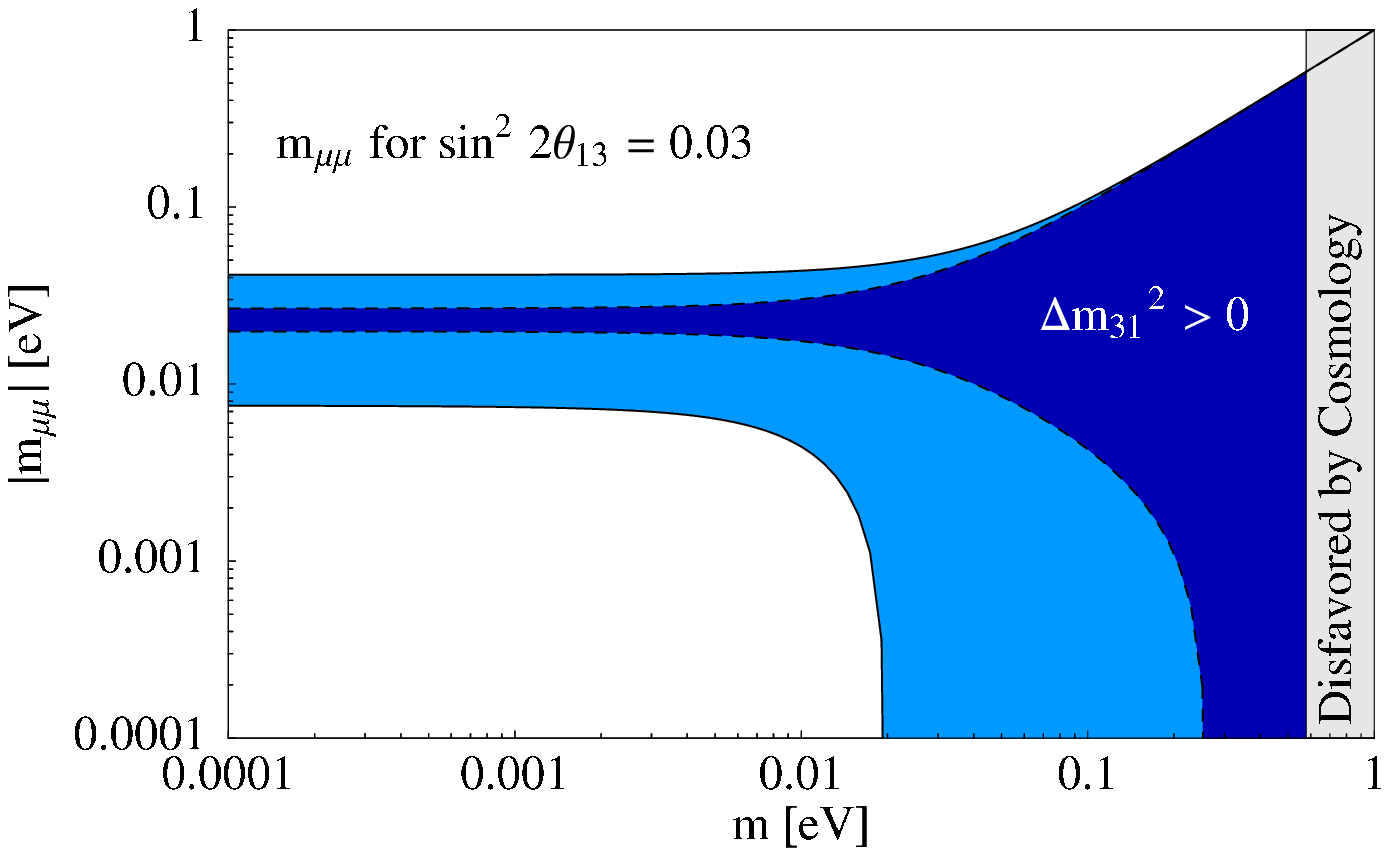,width=9.5cm,height=6.7cm} & \hspace{-1cm} 
\epsfig{file=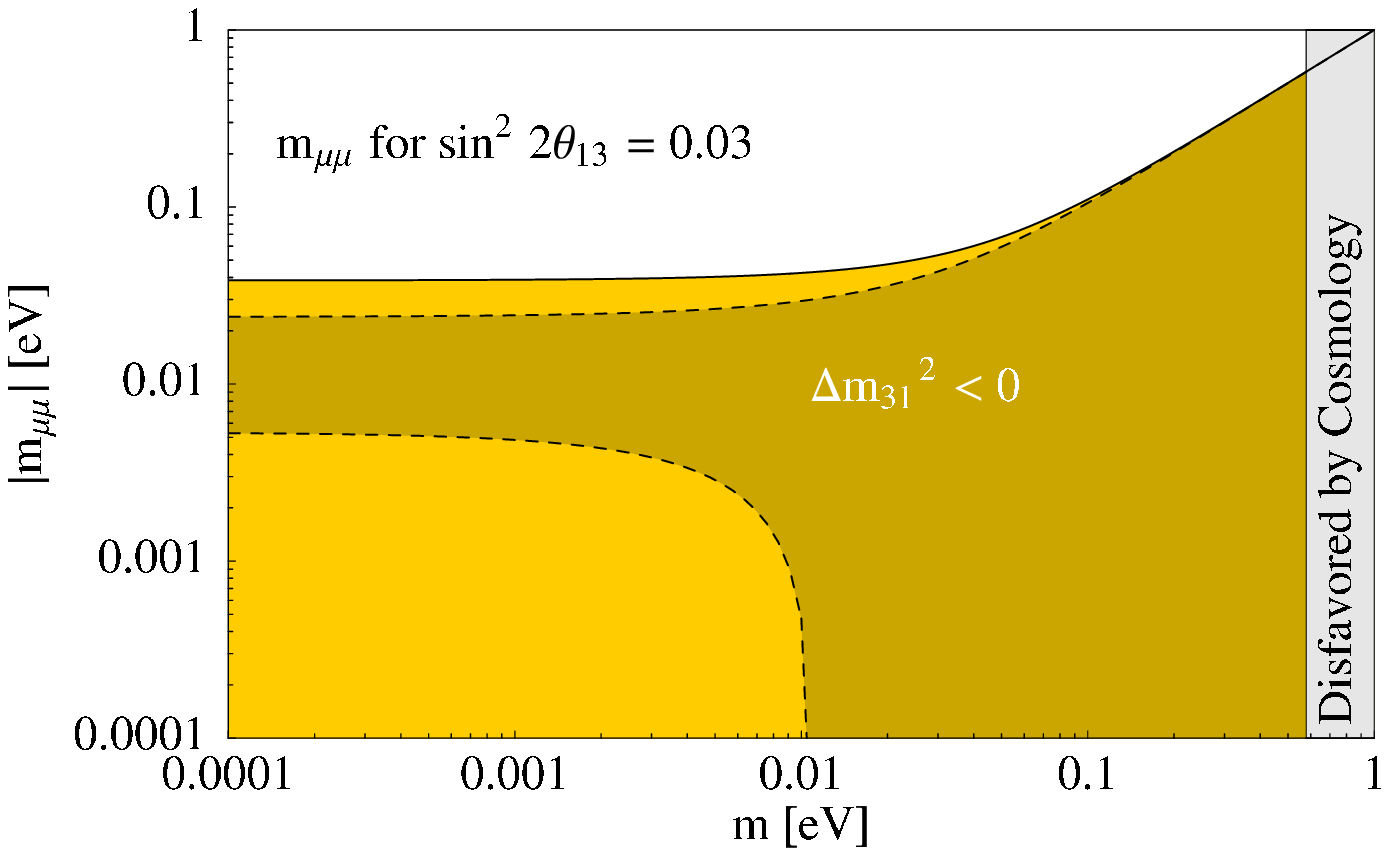,width=9.5cm,height=6.7cm} \\ \hspace{-1.2cm}
\epsfig{file=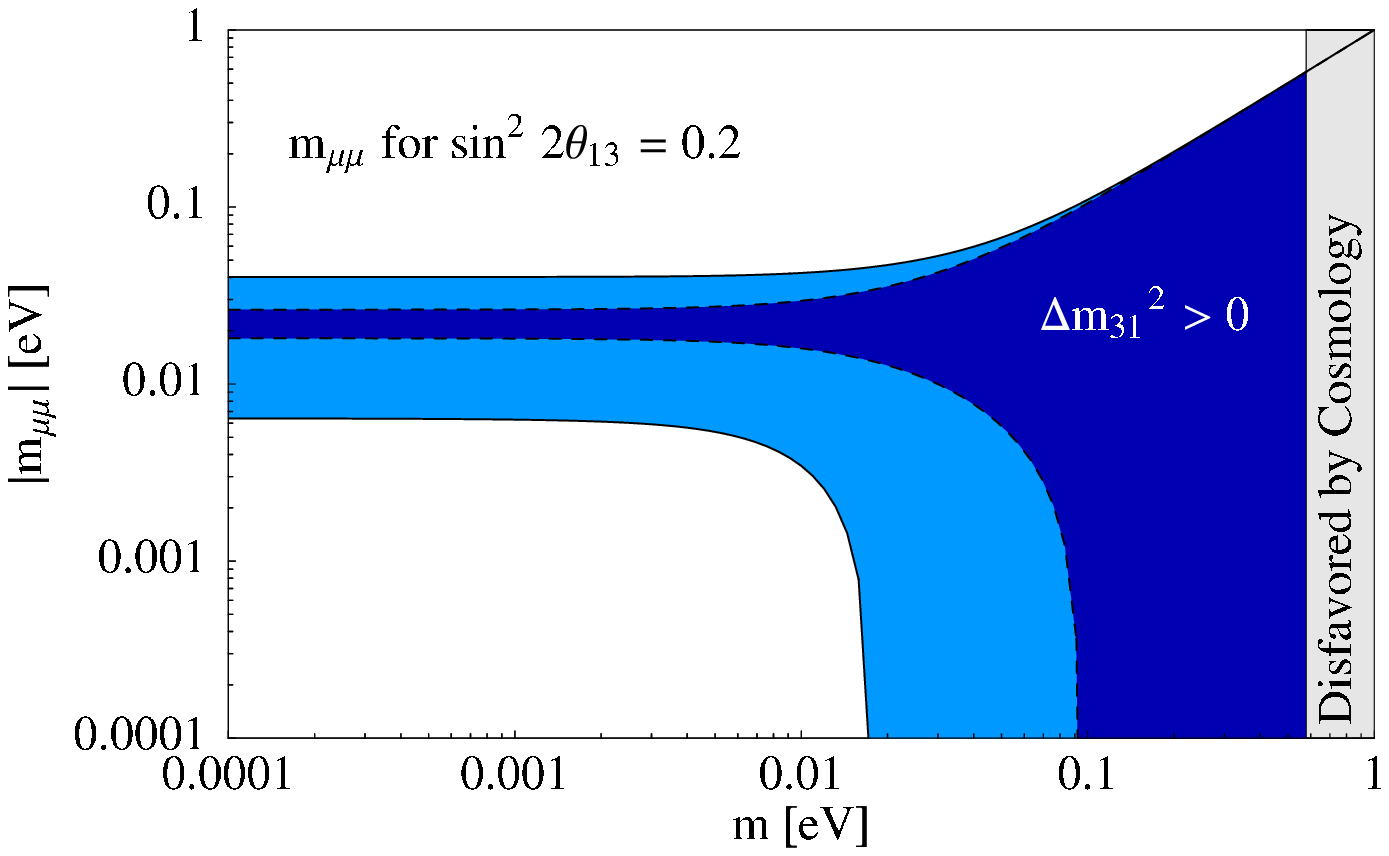,width=9.5cm,height=6.7cm} & \hspace{-1cm} 
\epsfig{file=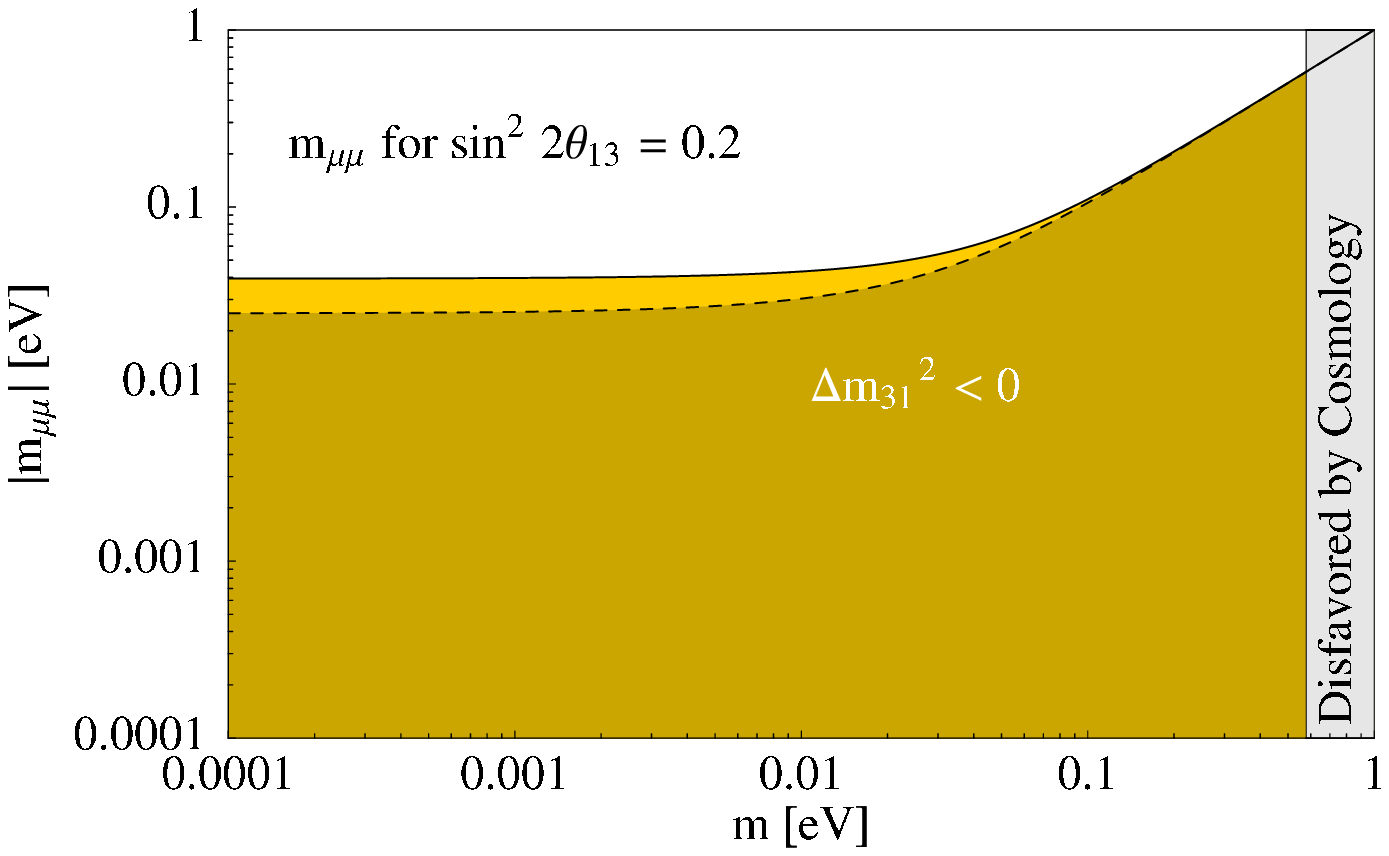,width=9.5cm,height=6.7cm}
\end{tabular}
\caption{\label{fig:mmm}The absolute value of the mass matrix 
element $m_{\mu\mu}$ against the 
smallest neutrino mass for the normal and inverted mass ordering for 
three representative values of $\theta_{13}$. The normal (inverted) mass 
ordering is given on the left (right) side. The 
best-fit and $3\sigma$ ranges of the oscillation parameters are used. 
The corresponding plot for $m_{\tau\tau}$ looks basically identical. 
}
\end{figure}

The $\mu\mu$ entry of the mass matrix reads 
\bea \label{eq:mmm}
m_{\mu \mu} = 
m_1\,{\left( c_{23}\,s_{12} +
        e^{i \delta}\,c_{12}\,s_{13}\,
         s_{23} \right) }^2 + 
 e^{2i\alpha}\,m_2\,
    {\left( c_{12}\,c_{23} -
        e^{i \delta}\,s_{12}\,s_{13}\,
         s_{23} \right) }^2  \\[0.3cm]
+ e^{2i\left( \beta + \delta \right) }\,
    m_3\,c_{13}^2\,s_{23}^2~.
\eea
In Fig.\ \ref{fig:mmm} we plot this element as a function of the smallest 
neutrino mass for both mass orderings and for the best-fit values 
and 3$\sigma$ ranges of the oscillation parameters. 

Let us again analyze the form of $m_{\mu\mu}$ in our two special cases:
\begin{itemize}
\item $\theta_{13}=0$: 
\bea \label{eq:mmm_t13}
(m_{\mu\mu})_{\theta_{13}=0} = c_{23}^2 
\left( e^{2i\alpha}\,m_2\, c_{12}^2 + m_1 \, s_{12}^2 \right)
+ e^{2i\left( \beta + \delta \right) }\,
    m_3\,s_{23}^2 \\[0.3cm]
\Rightarrow 
|m_{\mu \mu}|_{\theta_{13}=0} \simeq \left\{ 
\baz 
 s_{23} \sqrt{\dma \, s_{23}^2 + 2 \sqrt{\dms\, \dma} \,
c_{12}^2 \, c_{23}^2 \,c_{2( \alpha - \beta - \delta)}} 
& \mbox{ NH,} \\[0.2cm] 
\sqrt{\dma} \, c_{23}^2 \, \sqrt{1 - \sin^2 2\theta_{12} \, \sin^2\alpha} 
& \mbox{ IH,} \\[0.2cm] 
m_0 \left| 
c_{23}^2 \, \left( e^{2i\alpha} \, c_{12}^2 + s_{12}^2\right) 
+ e^{2i\left( \beta + \delta \right) }\,s_{23}^2
\right|
& \mbox{ QD.} 
\ea \right. 
\eea
The scale of $m_{\mu\mu}$ for zero $\theta_{13}$ and NH is 
roughly $\sqrt{\dma}/2 \simeq 0.02 $ eV. For IH it is approximately 
the same value, but cancellations up to 50 $\%$ are possible. 
For QD there can be complete cancellation, which we will discuss below.   
Numerically, for the best-fit values and the 1 and 3$\sigma$ ranges, 
we have 20.2--26.3 (15.8--31.1, 6.4--41.3) meV for NH, 
8.9--23.5 (6.6--27.9, 2.4--37.9) meV for IH. For QD with 
$m_0=0.5$ eV it holds 0.0--0.5 (0--0.5, 0--0.5) eV. 
Actually, there is a slight difference between normal and inverted ordering 
if neutrinos are quasi-degenerate in mass, see below.

For zero $\theta_{13}$ the element $m_{\mu\mu}$ 
cannot vanish for NH and IH. For NH and the best-fit values, the 
$\mu\mu$ entry has only a rather small range, centered around 
$\sqrt{\dma}/2$.  
The value of $m_{\mu\mu}$ is in case of 
IH nothing but the effective mass \meff~multiplied with $c_{23}^2$. 
If $\alpha=0$ then $m_{\mu\mu}$ takes its maximal value 
$ \sqrt{\dma} \, c_{23}^2 = c_{23}^2 \, \meff$.  
Moreover, as shown in Section \ref{sec:mem}, 
the $e\mu$ and $e\tau$ elements are very small in this case. 
If neutrinos are quasi-degenerate, Eq.~(\ref{eq:mmm_t13}) indicates that 
$m_{\mu\mu}$ vanishes if 
$c_{23}^2 \, \left( e^{2i\alpha} \, c_{12}^2 + s_{12}^2\right) 
+ e^{2i\left( \beta + \delta \right) }\,s_{23}^2$ vanishes. 
In case of $\theta_{23} = \pi/4$ this expression is zero for 
$\sin \alpha =0$ and $\sin( \beta + \delta) = 1$. 
Looking at Fig.~\ref{fig:mmm}, however, one sees that for 
vanishing $\theta_{13}$ the curve for minimal $|m_{\mu\mu}|$ and normal 
ordering does not 
coincide with the one for inverted ordering, and in addition is non-zero. 
To understand this, let us 
define $2 \, r_\odot \equiv \dms/m_0^2$ and 
$2 \, r_{\rm A} \equiv \dma/m_0^2$, which allows from 
Eq.\ (\ref{eq:masses}) to write 
$m_1 = m_0$, $m_3 \simeq m_0 \, (1 + r_{\rm A})$ and 
$m_2 \simeq m_0 \, (1 + r_{\rm \odot})$ for the normal ordering and 
$m_3 = m_0$, $m_1 \simeq m_0 \, (1 + r_{\rm A})$ and 
$m_2 \simeq m_0 \, (1 + r_{\rm A} + r_{\rm \odot})$ for the 
inverted ordering. Within this approximation, the formulae for 
$m_{\mu\mu}$ in case of normal and inverted ordering read: 
\be \label{eq:NHvsIH}\hspace{-0.51cm}
\left(m_{\mu\mu}\right)_{\scriptsize \theta_{23}=\pi/4,\,\theta_{13}=0}
^{\rm QD} \simeq 
\left\{ 
\baz  
\frac 12 m_0  \left( s_{12}^2 + c_{12}^2 \, e^{2i\alpha} \, (1+r_{\odot}) 
 + e^{2i(\beta+\delta)} \, (1+r_{\rm A})\right) & 
\mbox{ normal,} \\[0.2cm] 
\frac 12 m_0  \left( s_{12}^2 \, (1+r_{\rm A}) + 
c_{12}^2 \, e^{2i\alpha} \, (1+r_{\rm A}+r_{\odot}) + e^{2i(\beta+\delta)} 
\right) & \mbox{ inverted.} 
\ea \right. 
\ee
For both possibilities, the term proportional to 
 $e^{2i(\beta+\delta)}$ dominates the other two. However, for the 
inverted ordering the sum of the two remaining terms can exceed the 
contribution from the third, leading term. Hence, the 
$\mu\mu$ entry can vanish for the inverted ordering. 
In case of normal ordering, however, the sum of the two sub-leading 
terms is always 
smaller than the third, leading term and therefore the minimal value of the 
$\mu\mu$ element is non-zero. For $m_0 = 0.3$ eV it is given by 
$\frac 12 m_0 \, ( (1 + r_{\rm A}) - c_{12}^2 \, (1+r_{\rm A}+r_{\odot}) 
- s_{12}^2 ) =  \frac 12 m_0 \, (r_{\rm A} - c_{12}^2 \, r_\odot) \simeq 
0.0018$ eV. 
This agrees well with the value in Fig.\ \ref{fig:mmm}. 
Note also that from Eq.~(\ref{eq:NHvsIH}) one finds that 
the minimal values of $m_{\mu\mu}$ are different for normal and 
inverted ordering, with this difference given also by 
$\frac 12 m_0 \, (r_{\rm A} - c_{12}^2 \, r_\odot)$. 
This is in contrast to $m_{ee}$, 
$m_{e \mu}$ and $m_{e \tau}$, which show no such difference for 
neutrino masses larger than 0.1 eV. It is easy to show that, for $\theta_{13}=0$, the difference
between the 
minimal value of $m_{ee}$ for normal and inverted ordering is 
of order $m_0 \, r_{\rm A}$. However, since the minimal value of \meff~is 
much larger than this difference 
(namely of the order $m_0 \, \cos 2 \theta_{12}$), 
the difference cannot be seen in Fig.~\ref{fig:mee}. 
For $m_{e \mu}$ and $m_{e \tau}$ it turns out that the difference 
between the minimal values  for normal and inverted ordering vanishes. 
 
Nevertheless, the approximate expression for QD in 
Eq.\ (\ref{eq:mmm_t13}) indicates the presence of an 
interesting correlation between $\alpha$ and $\theta_{23}$, obtained when 
one requires the $\mu\mu$ entry to vanish. At the end of this Section we 
will study this in detail.

\item $\theta_{23}=\pi/4$: 
\bea \label{eq:mmm_t23}\hspace{-1cm}
(m_{\mu \mu})_{\theta_{23}=\pi/4} = \frac{1}{2} \left( 
m_1\,{\left( s_{12} +
        e^{i \delta}\,c_{12}\,s_{13}\,
          \right) }^2 + 
 e^{2i\alpha}\,m_2\,
    {\left( c_{12} -
        e^{i \delta}\,s_{12}\,s_{13}\,
          \right) }^2 + 
e^{2i\left( \beta + \delta \right) }\,
    m_3\,c_{13}^2 \right) \\[0.3cm]\hspace{-0.6cm}
\Rightarrow 
|m_{\mu \mu}|_{\theta_{23}=\pi/4} \simeq \left\{ 
\baz 
\frac 12 \, c_{13}^2 \,  
\sqrt{\dma \, c_{13}^2 + 2 \sqrt{\dms\, \dma} \,
c_{12}^2  \,c_{2(\alpha - \beta - \delta)}} 
& \mbox{ NH,} \\[0.2cm] 
\frac 12 \, \sqrt{\dma} \, \left| 
e^{2 i \alpha} \, c_{12}^2 + s_{12}^2 + 2 \, e^{i \delta} 
\left(1 - e^{2 i \alpha} \right) \, c_{12} \, s_{12} \, s_{13} 
\right|
& \mbox{ IH,} \\[0.2cm] 
\frac 12 \, m_0 \, \left| 
e^{2 i \alpha} \, c_{12}^2 + s_{12}^2 
+ e^{2i\left( \beta + \delta \right) }\,c_{13}^2
+ 2 \, e^{i \delta} 
\left(1 - e^{2 i \alpha} \right) \, c_{12} \, s_{12} \, s_{13} 
\right|
& \mbox{ QD.} 
\ea \right. 
\eea
Numerically, one gets 0.9--1.2 (0.8--1.4, 0.1--1.8) meV for NH, 
8.9--23.5 (2.9--25.9, 0--31.3) meV for IH and for QD 
with $m_0=0.5$ eV we have 0--0.50 eV (0--0.50 eV, 0--0.51 eV). 
Comparing the expressions with the case of vanishing $\theta_{13}$ in 
Eq.\ (\ref{eq:mmm_t13}) we see that the corrections 
are rather small. One notable difference is that $m_{\mu\mu}$ can vanish 
now for IH. 
\end{itemize}

Again, the (approximate) $\mu$-$\tau$ symmetry implies that 
$m_{\mu\mu} \simeq m_{\tau\tau}$. 
The explicit form of $m_{\tau\tau}$ is 
\bea \label{eq:mtt} 
m_{\tau\tau} = 
m_1\,{\left( 
        e^{i \delta}\,c_{12}\,s_{13}\,
         c_{23} - s_{23}\,s_{12}  \right) }^2 + 
 e^{2i\alpha}\,m_2\,
    {\left( c_{12}\,s_{23} + 
        e^{i \delta}\,s_{12}\,s_{13}\,
         c_{23} \right) }^2 \\[0.3cm]
+ 
e^{2i\left( \beta + \delta \right) }\,
    m_3\,c_{13}^2\,c_{23}^2~.
\eea
It is obtained from $m_{\mu \mu}$ by exchanging $s_{23}$ with $c_{23}$ 
and $c_{23}$ with $-s_{23}$. 

\begin{figure}[tb]
\hspace{-1.2cm}
\begin{tabular}[h]{lr}\hspace{-1.2cm}
\epsfig{file=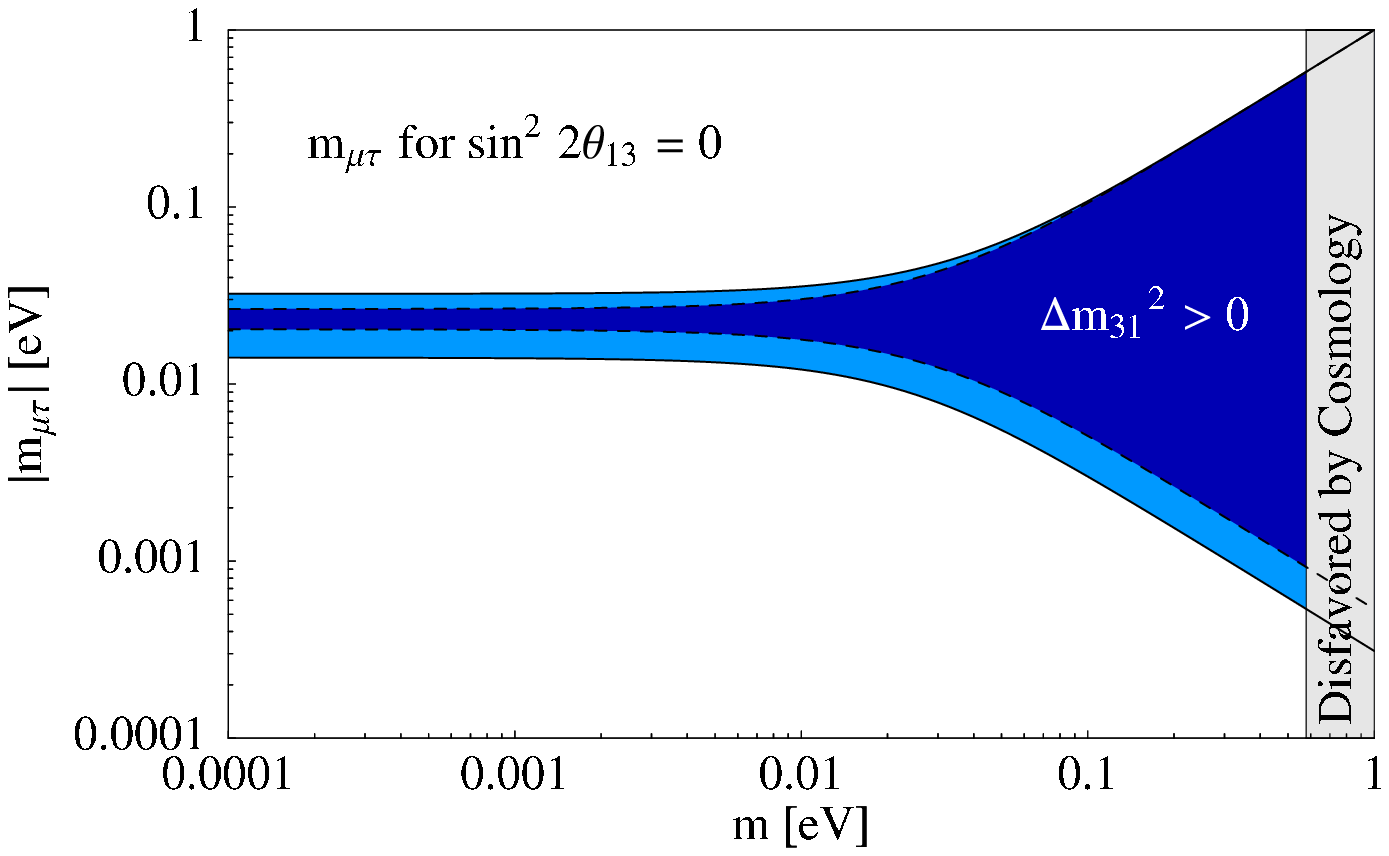,width=9.5cm,height=6.7cm} & \hspace{-1cm}  
\epsfig{file=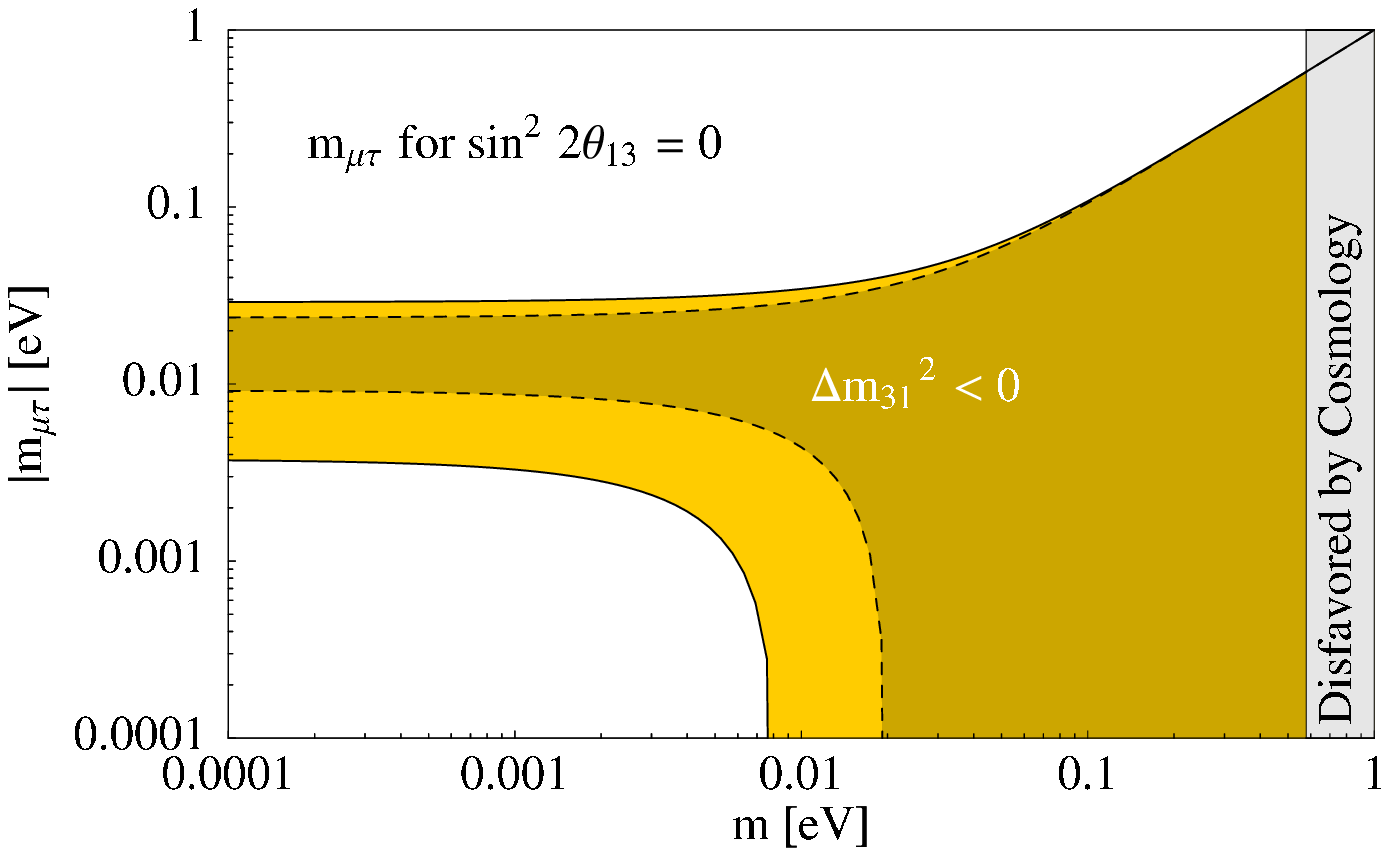,width=9.5cm,height=6.7cm}\\ \hspace{-1.2cm}
\epsfig{file=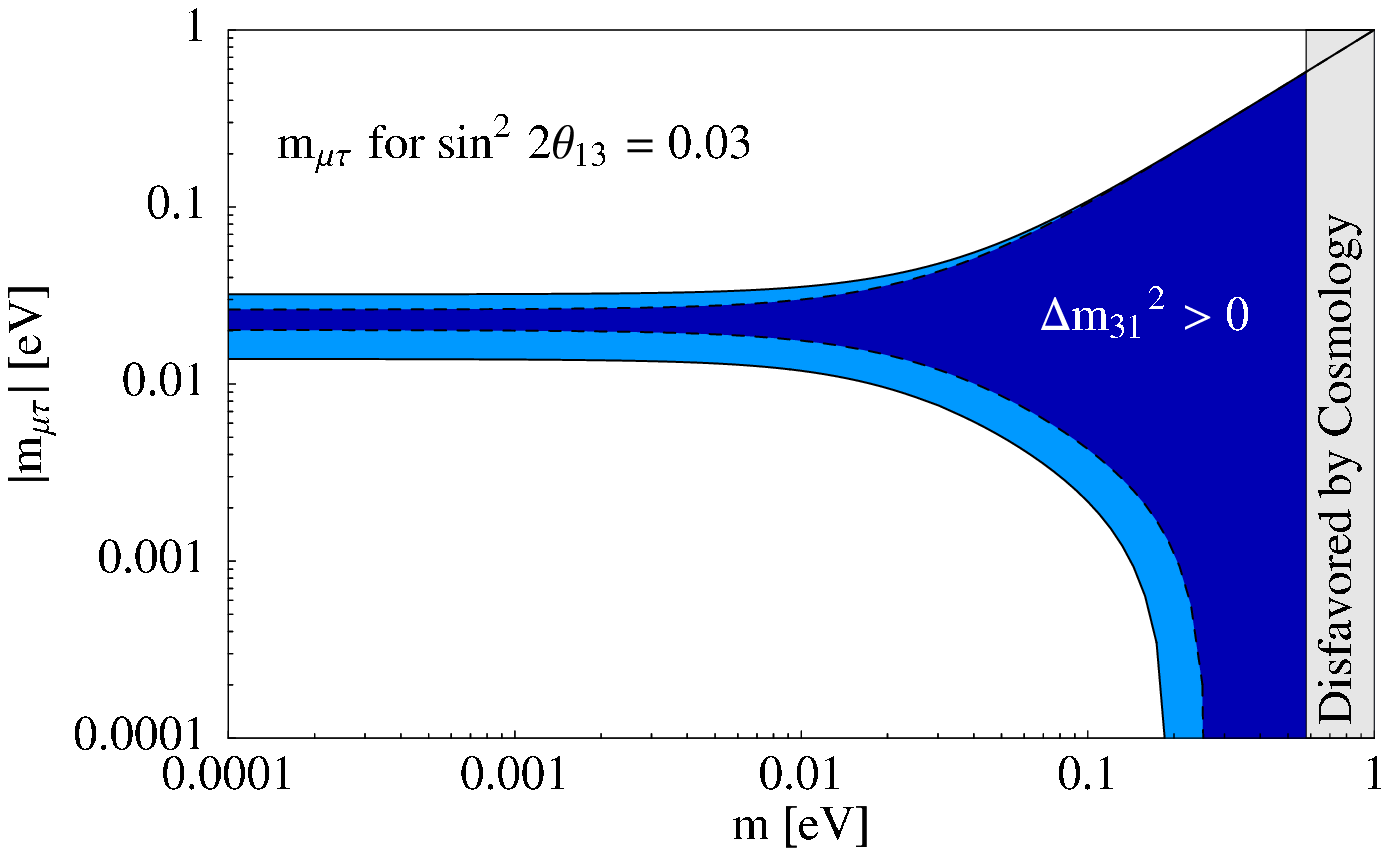,width=9.5cm,height=6.7cm} & \hspace{-1cm} 
\epsfig{file=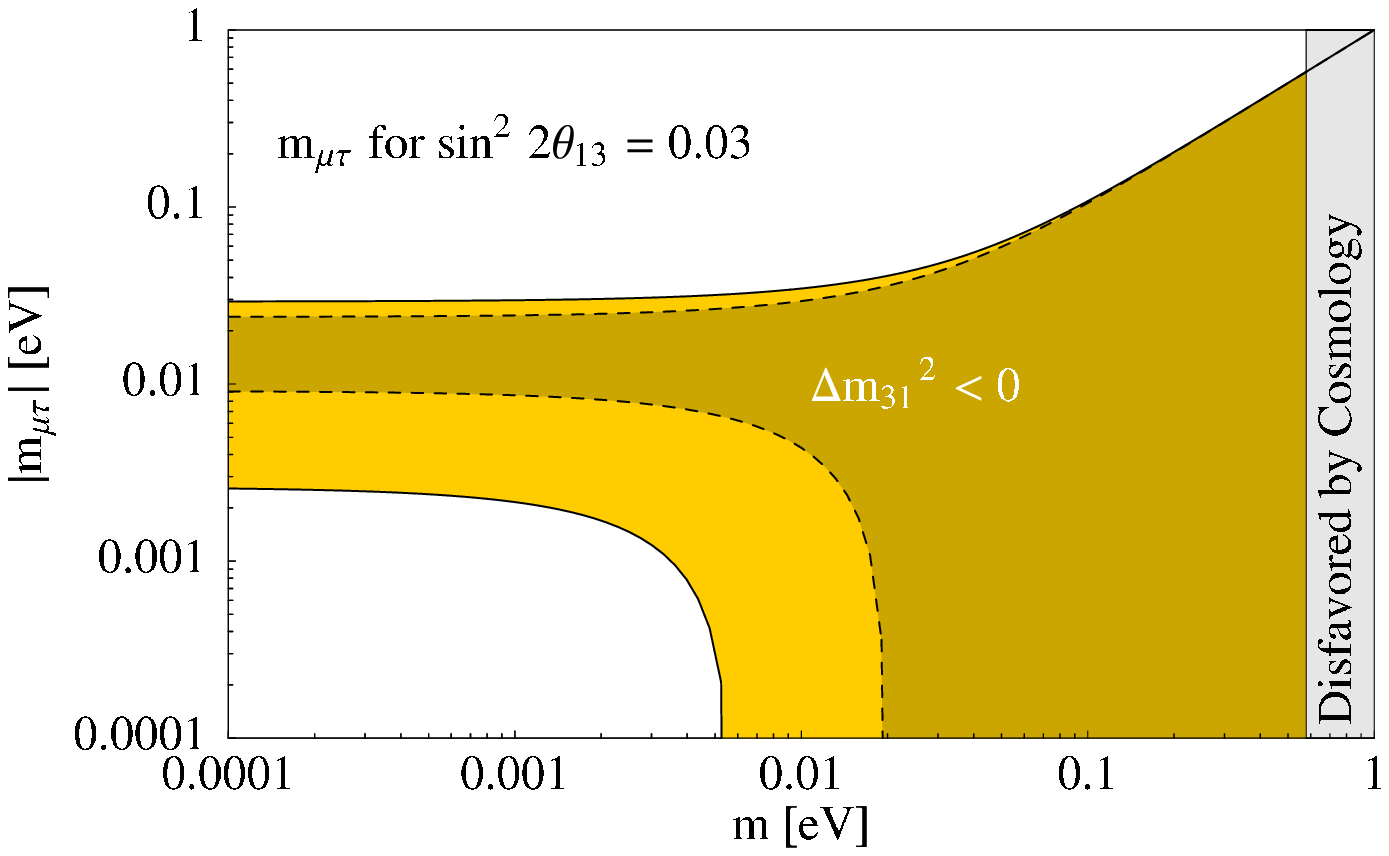,width=9.5cm,height=6.7cm}\\ \hspace{-1.2cm}
\epsfig{file=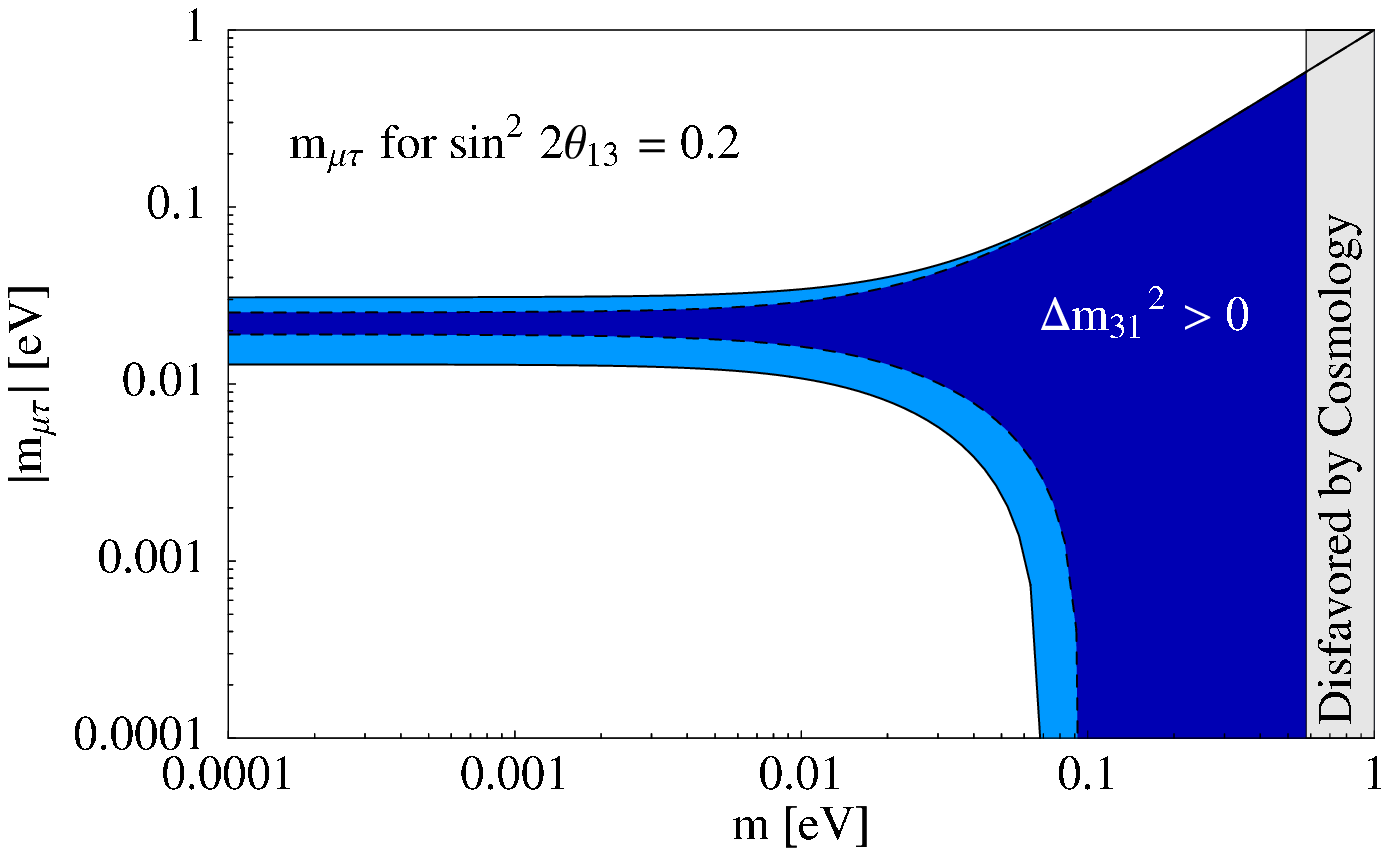,width=9.5cm,height=6.7cm} & \hspace{-1cm} 
\epsfig{file=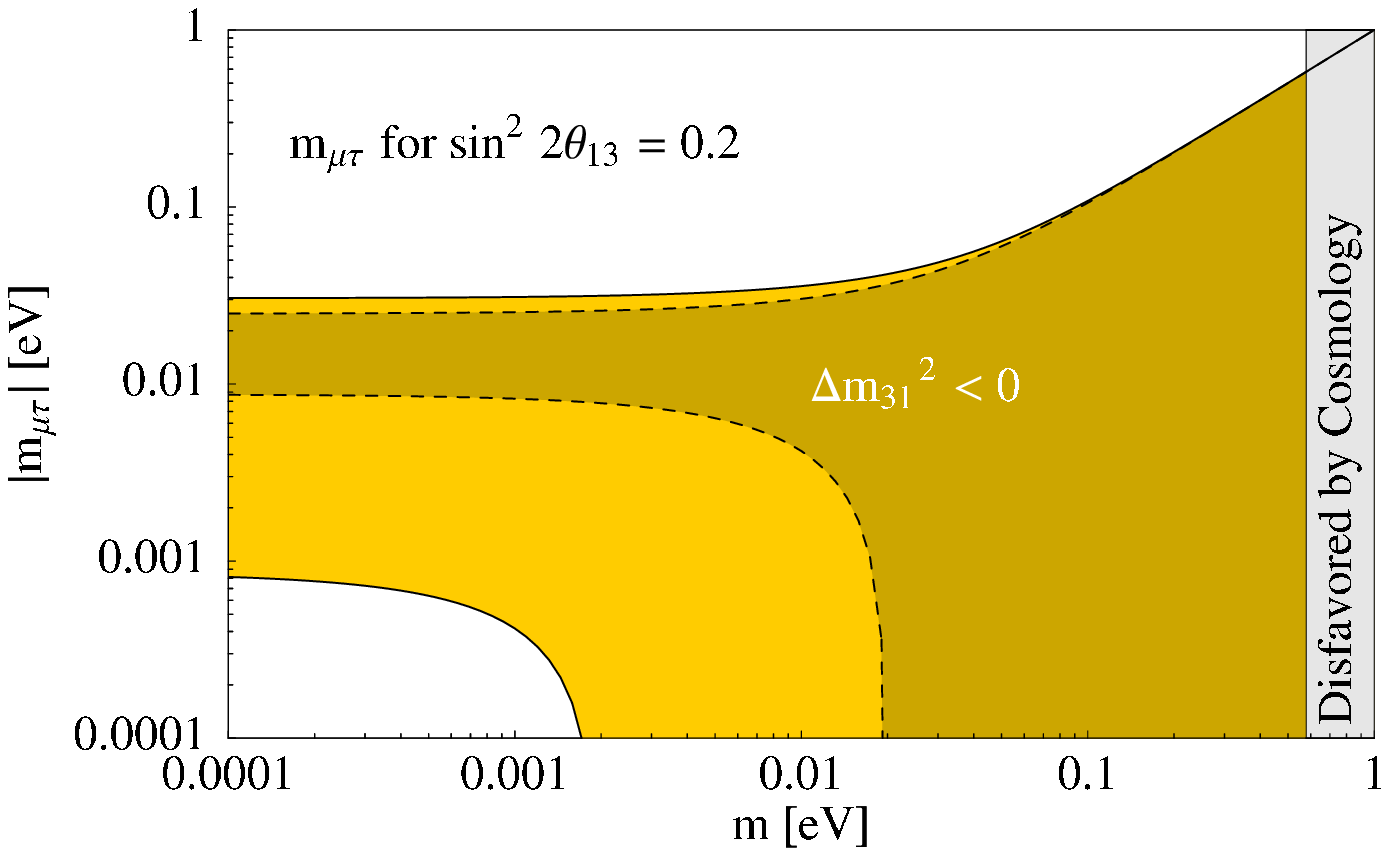,width=9.5cm,height=6.7cm}
\end{tabular}
\caption{\label{fig:mmt}The absolute value of the mass matrix 
element $m_{\mu\tau}$ against the 
smallest neutrino mass for the normal and inverted mass ordering for 
three representative values of $\theta_{13}$. The normal (inverted) mass 
ordering is given on the left (right) side. The 
best-fit and $3\sigma$ ranges of the oscillation parameters are used.
}
\end{figure}

The mass matrix element $m_{\mu \tau}$ reads:
\bea \label{eq:mmt} 
|m_{\mu\tau}| = 
\frac 12  \,
      \cos 2\theta_{23}\, 
\, e^{i\delta}\,\left( m_1 -
        e^{2i \alpha}\,m_2 \right) \sin 2\theta_{12}\,
      s_{13}- \\[0.3cm] 
\frac 12 \, \sin 2 \theta_{23} 
\, \left( e^{2i\alpha}\,m_2\,c_{12}^2  + 
      m_1\,s_{12}^2 - 
      e^{2i \delta}\,
       \left( e^{2i \beta}\,m_3\,
          c_{13}^2 +
         \left( m_1\,c_{12}^2 +
            e^{2i  \alpha}\,m_2\,
             s_{12}^2 \right) \,s_{13}^2
         \right)  \right) ~.
\eea
The first term is suppressed by the close-to-maximal $\theta_{23}$ and the 
smallness of $\theta_{13}$. 
One can show that $m_{\mu\tau}$ 
is obtained from $m_{\mu \mu}$ by exchanging $c_{23}^2$ with 
$-c_{23} \, s_{23}$, $s_{23}^2$ with $c_{23} \, s_{23}$ and 
$c_{23} \, s_{23}$ with $1/2\, (c_{23}^2 - s_{23}^2)$. 
A plot of $|m_{\mu \tau}|$ as a function of the smallest mass 
is given in Fig.\ \ref{fig:mmt} and looks rather similar to the 
cases of $|m_{\mu \mu}|$ and $|m_{\tau \tau}|$. 
Note that for $\theta_{13}=0$ 
the minimal values for the normal and inverted ordering show differences. 
Moreover, for the normal mass ordering and quasi-degenerate masses 
$m_{\mu \tau}$ cannot vanish when $\theta_{13}=0$. 
This can be explained in analogy to the issues discussed above 
for $m_{\mu\mu}$. 
Non-zero $\theta_{13}$ and normal ordering 
allow for zero $m_{\mu\tau}$ only in the QD regime. In case of IH and NH, 
the $\mu\tau$ entry cannot vanish.\\

\begin{figure}[tb]
\begin{center}
\epsfig{file=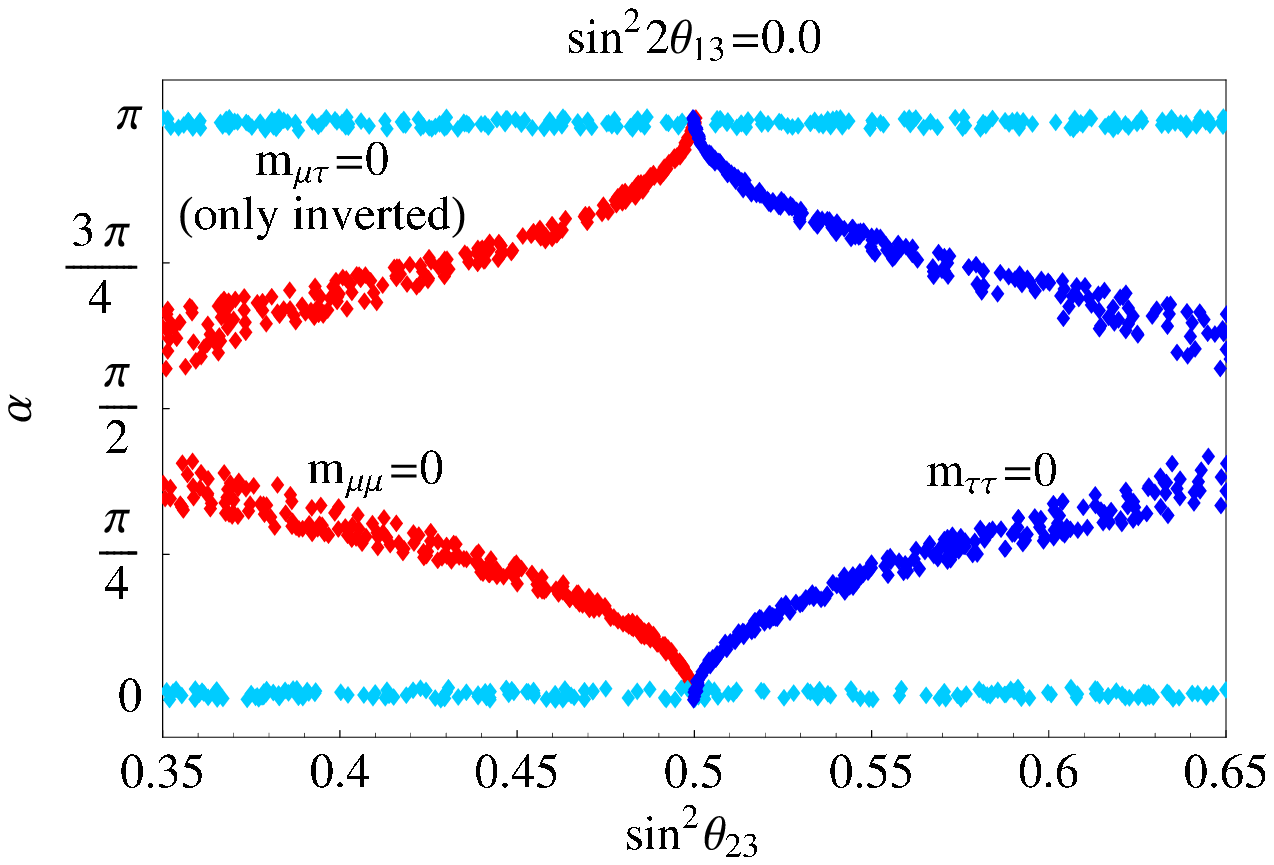,width=8cm,height=6cm}
\epsfig{file=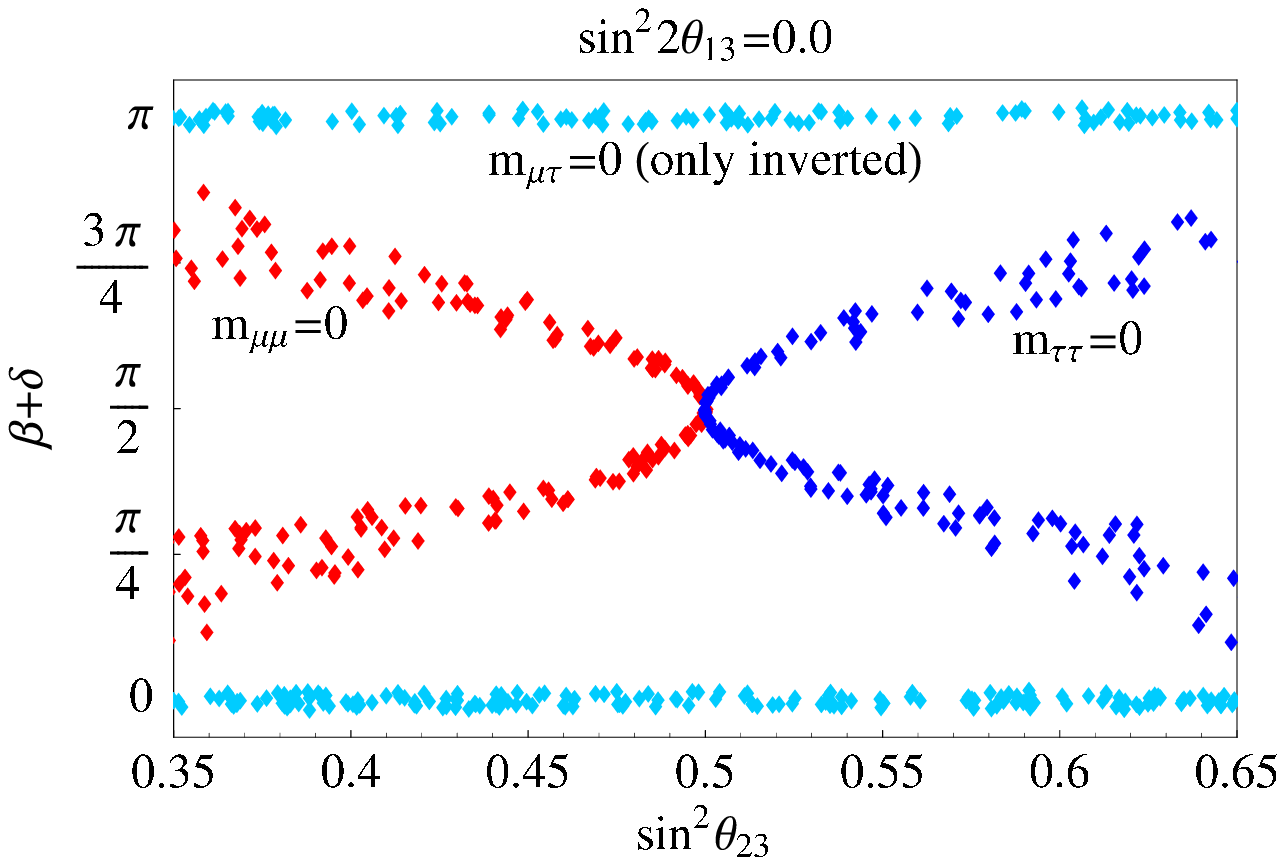,width=8cm,height=6cm}
\epsfig{file=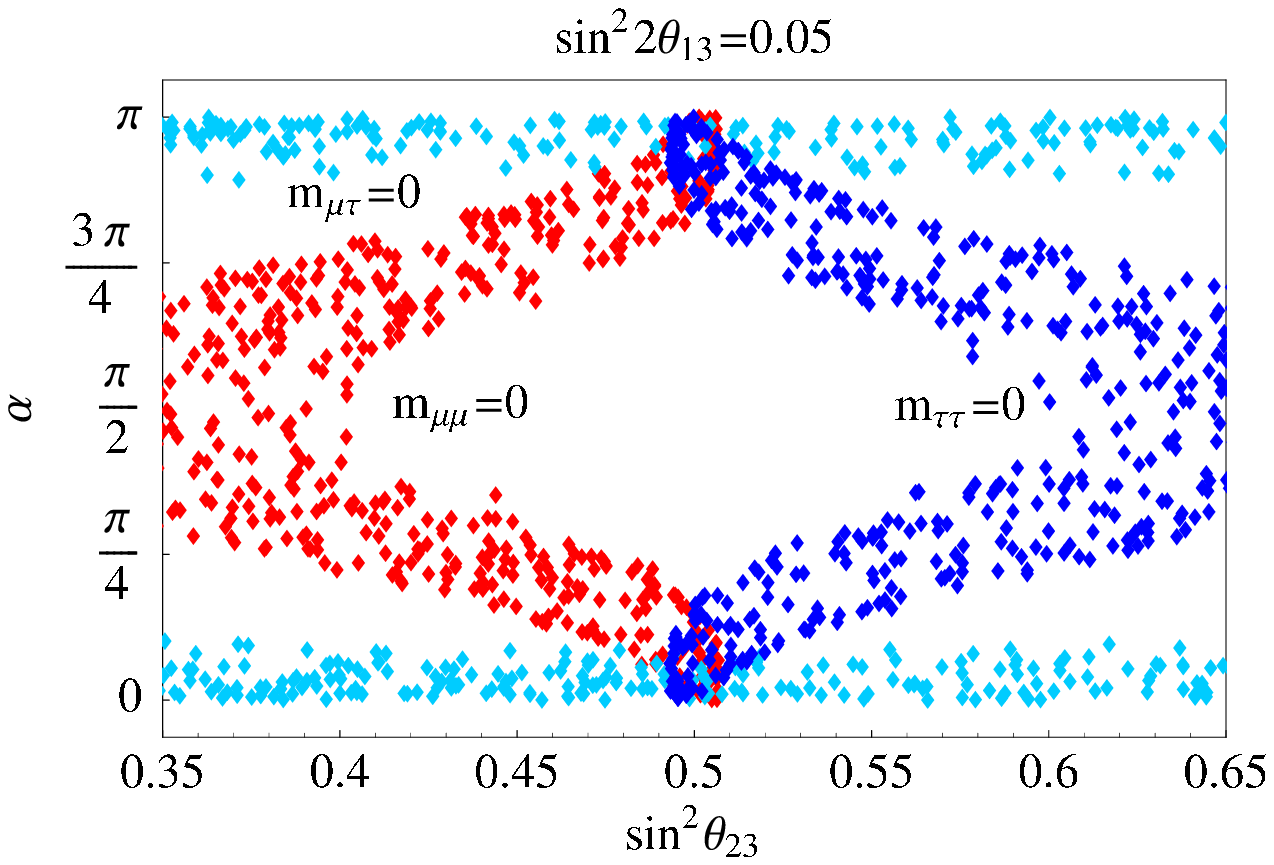,width=8cm,height=6cm}
\epsfig{file=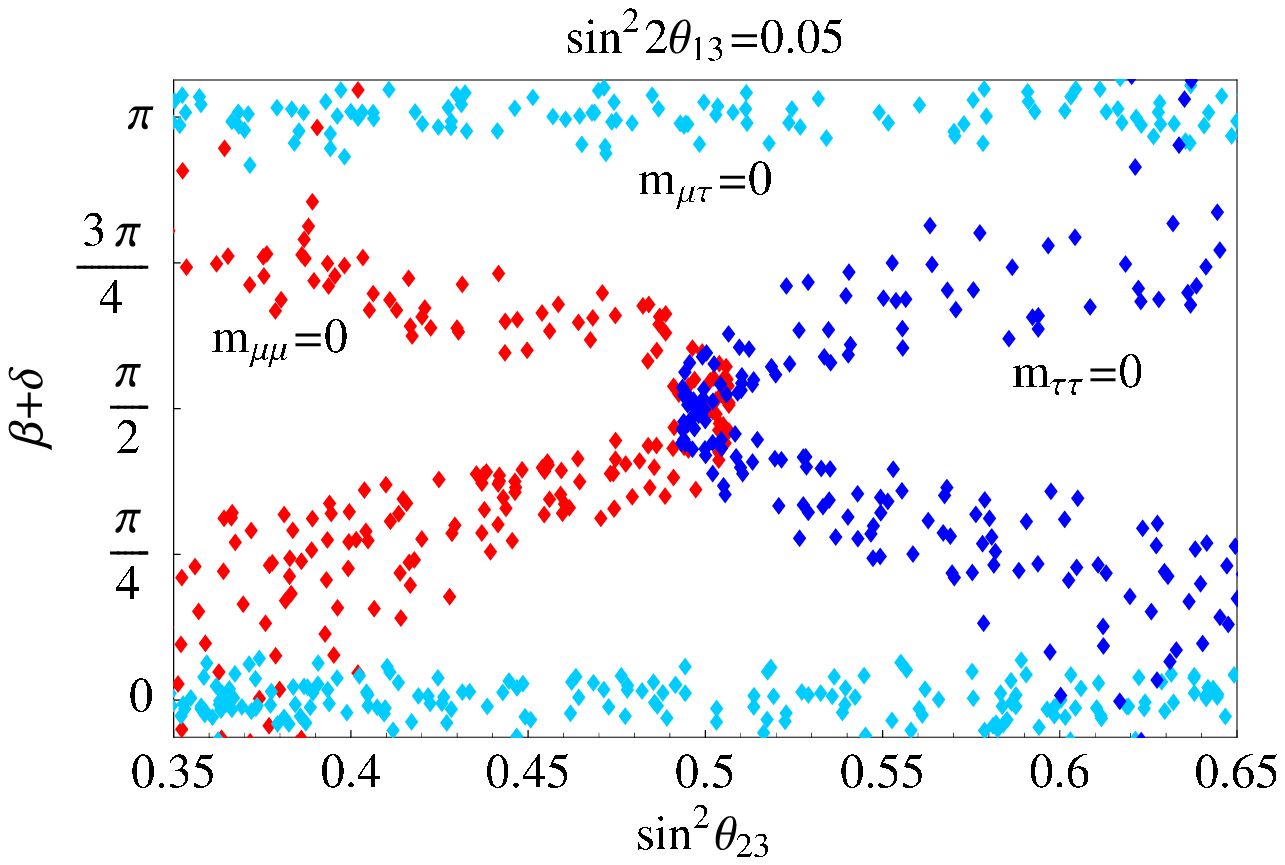,width=8cm,height=6cm}
\caption{\label{fig:mmm1}QD neutrinos, vanishing 
$\mu\mu$, $\tau\tau$ or $\mu\tau$ entries and the 
resulting correlation between $\theta_{23}$ and $\alpha$, 
and between $\theta_{23}$ and $\beta + \delta$.}
\end{center}
\end{figure}

It is clear from the above discussion that an 
interesting correlation in case of a vanishing 
$\mu\mu$, $\mu\tau$ or $\tau\tau$ element can occur only 
if neutrinos are quasi-degenerate. Let us summarize 
the form of the three relevant mass matrix elements for QD: 
\bea \label{eq:mtblock}
|m_{\mu\mu}|^{\rm QD} \simeq m_0 
\left| 
c_{23}^2 \, \left( e^{2i\alpha} \, c_{12}^2 + s_{12}^2\right) 
+ e^{2i\left( \beta + \delta \right) }\,s_{23}^2
\right|~, \\[0.3cm]
|m_{\tau\tau}|^{\rm QD} \simeq m_0 
\left| 
s_{23}^2 \, \left( e^{2i\alpha} \, c_{12}^2 + s_{12}^2\right) 
+ e^{2i\left( \beta + \delta \right) }\,c_{23}^2
\right|~, \\[0.3cm]
|m_{\mu\tau}|^{\rm QD} \simeq m_0 \, c_{23} \, s_{23} \, 
\left| 
\left( e^{2i\alpha} \, c_{12}^2 + s_{12}^2\right) 
- e^{2i\left( \beta + \delta \right) } 
\right| ~.
\eea
Recall that for $\theta_{13}=0$ the $\mu\tau$ element cannot vanish if 
neutrinos are normally ordered. 
Including finite values of $\theta_{13}$ will lead 
to modifications of the order $\sin \theta_{13}$ in the above formulae. 
It is apparent that correlations between the phases and 
$\theta_{23}$ are implied. 
For the $\mu\mu$ entry we can immediately see from Eq.~(\ref{eq:mtblock}) 
that for $\theta_{23} > \pi/4$ the mass matrix element 
cannot vanish. Consequently, a vanishing $\tau\tau$ element is not 
possible for $\theta_{23} < \pi/4$. However, finite values of $\theta_{13}$ 
will allow for values of $\theta_{23}$ slightly below (above) $\pi/4$. 
Exactly maximal $\theta_{23}=\pi/4$ will lead to vanishing 
$\mu\mu$ or $\tau\tau$ entries if 
$\sin \alpha = 0$ and $\sin(\beta + \delta) = 1$. 
For the $\mu\tau$ entry there is no dependence on $\theta_{23}$, 
and $m_{\mu\tau}=0$ is only possible for values of the phases 
corresponding to $\sin \alpha = \sin(\beta + \delta) = 0$. 
In Fig.~\ref{fig:mmm1} we show for two values of $\theta_{13}$ scatter 
plots of $\alpha$ against $\sin^2 \theta_{23}$ and of 
$\beta + \delta$ against $\sin^2 \theta_{23}$, 
confirming these considerations. We choose two values of 
$\theta_{13}$ and varied the remaining oscillation parameters within 
their $3\sigma$ ranges.

\begin{figure}[tb]
\begin{center}
\epsfig{file=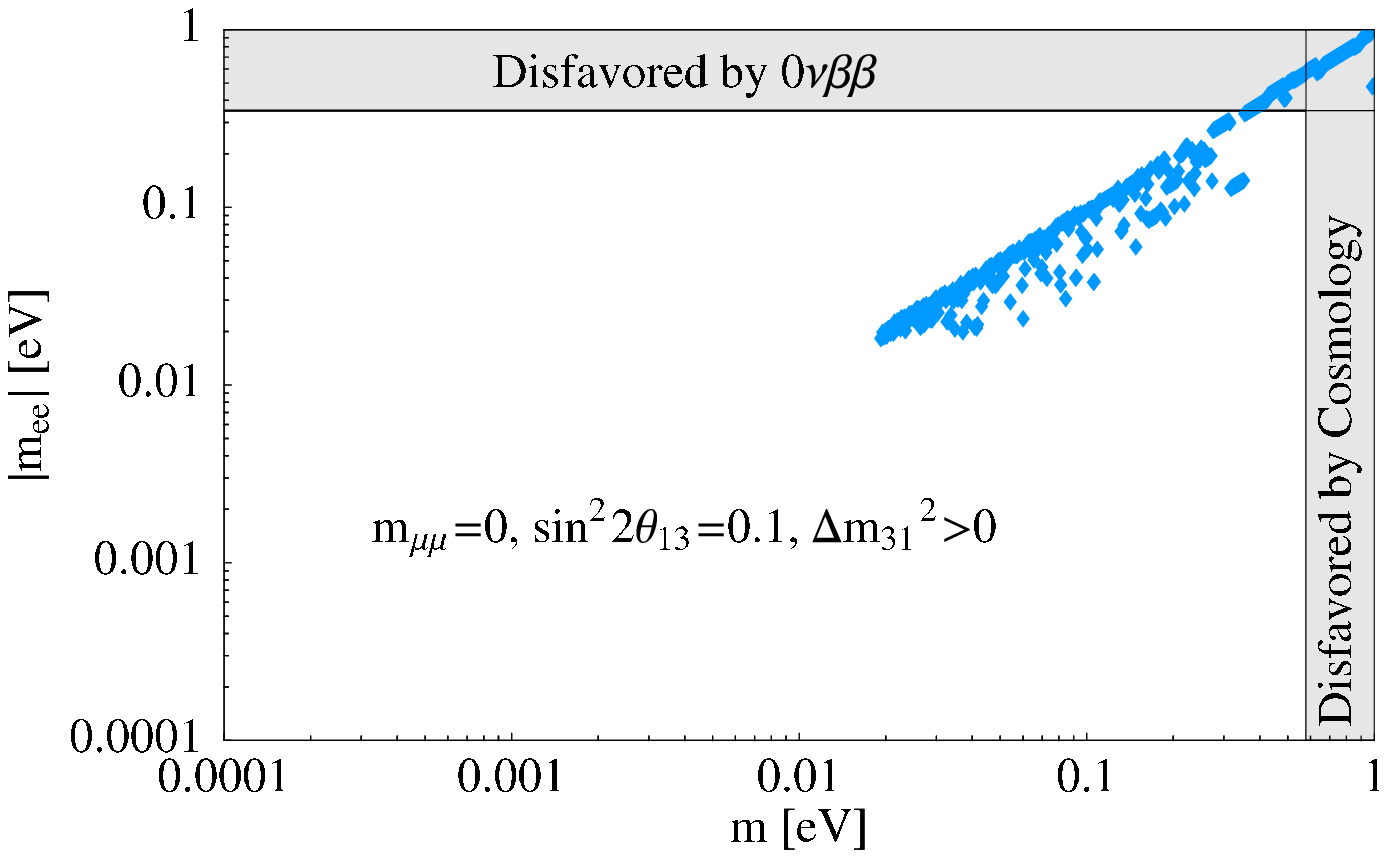,width=8cm,height=6cm}
\epsfig{file=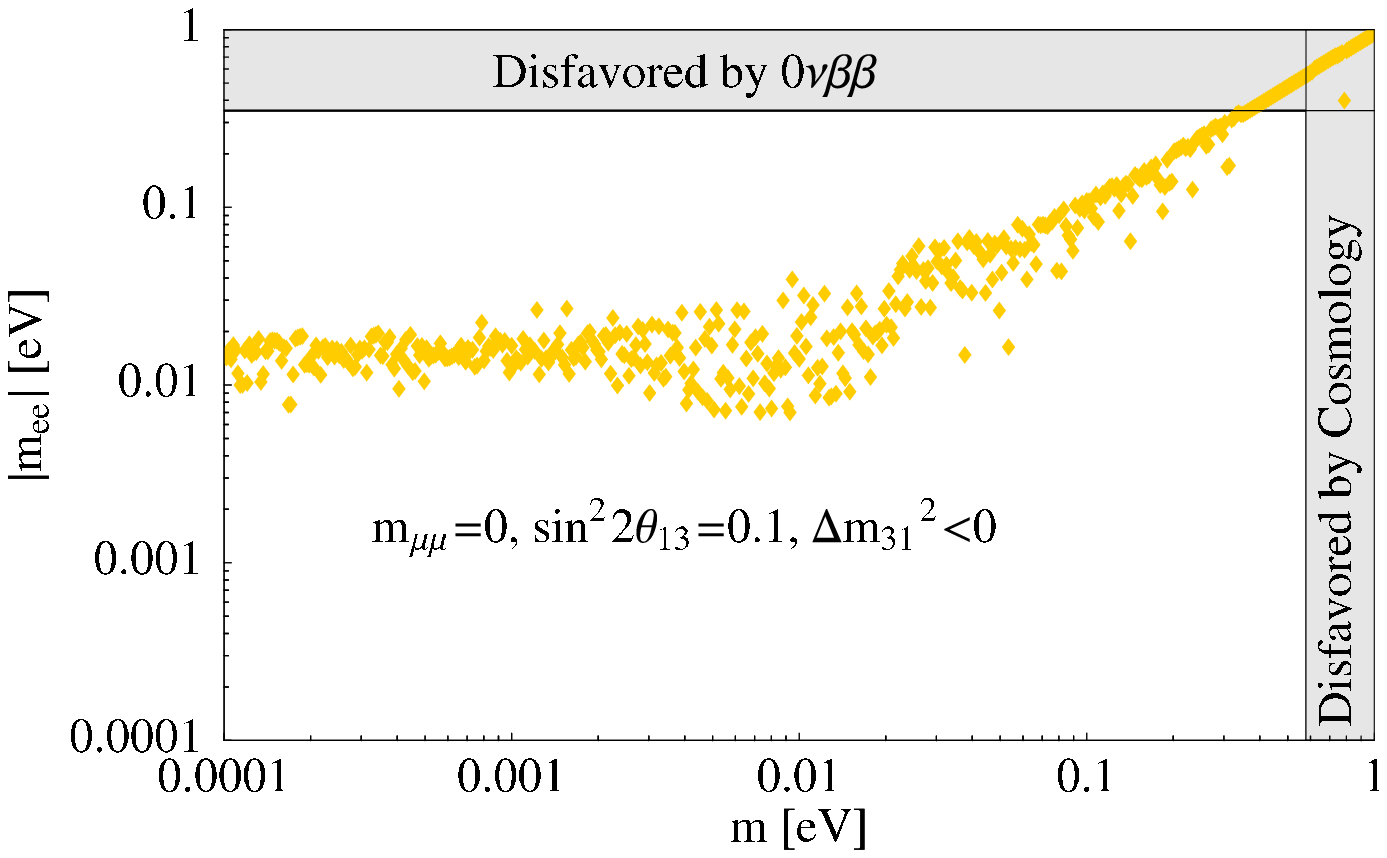,width=8cm,height=6cm}
\epsfig{file=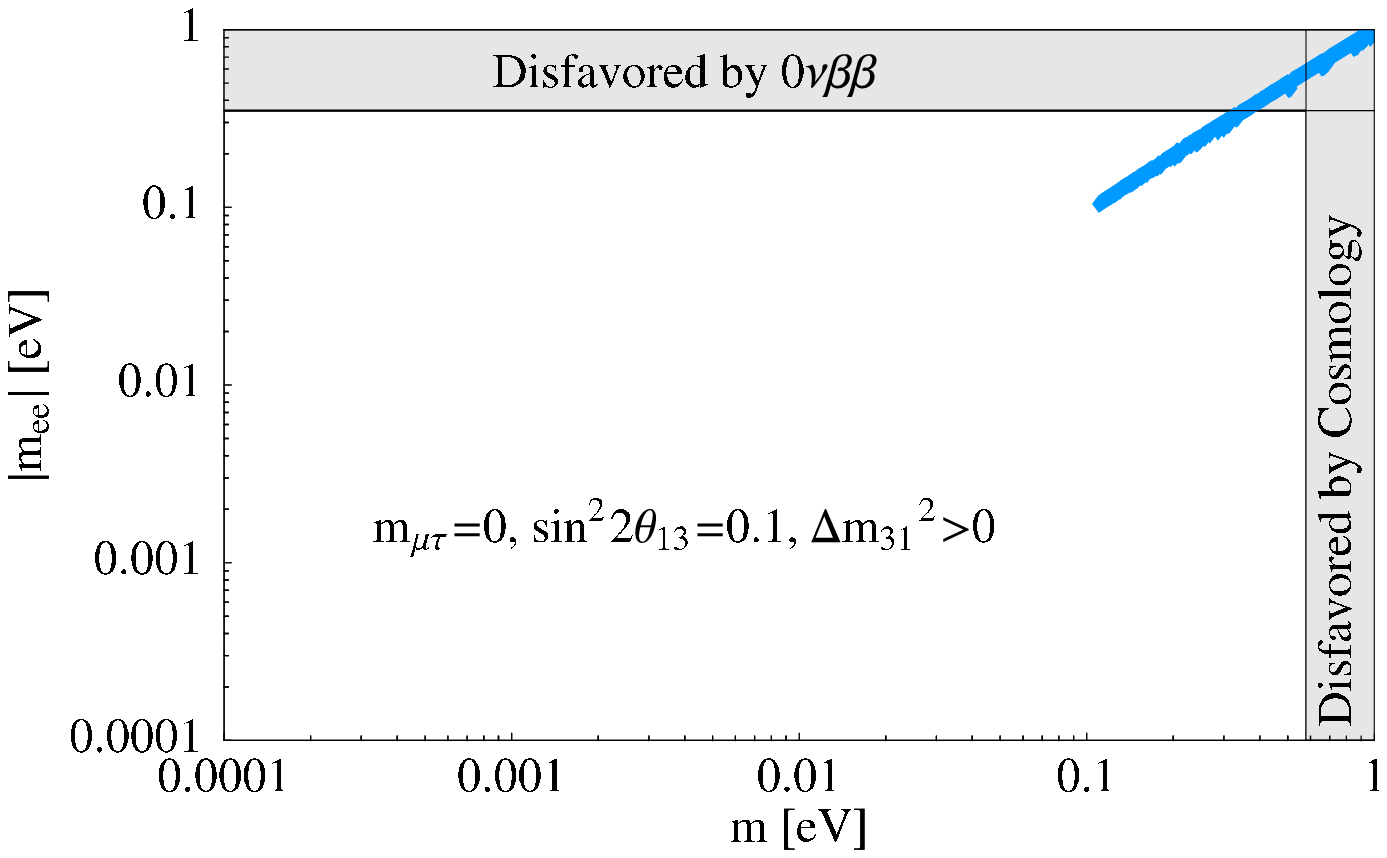,width=8cm,height=6cm}
\epsfig{file=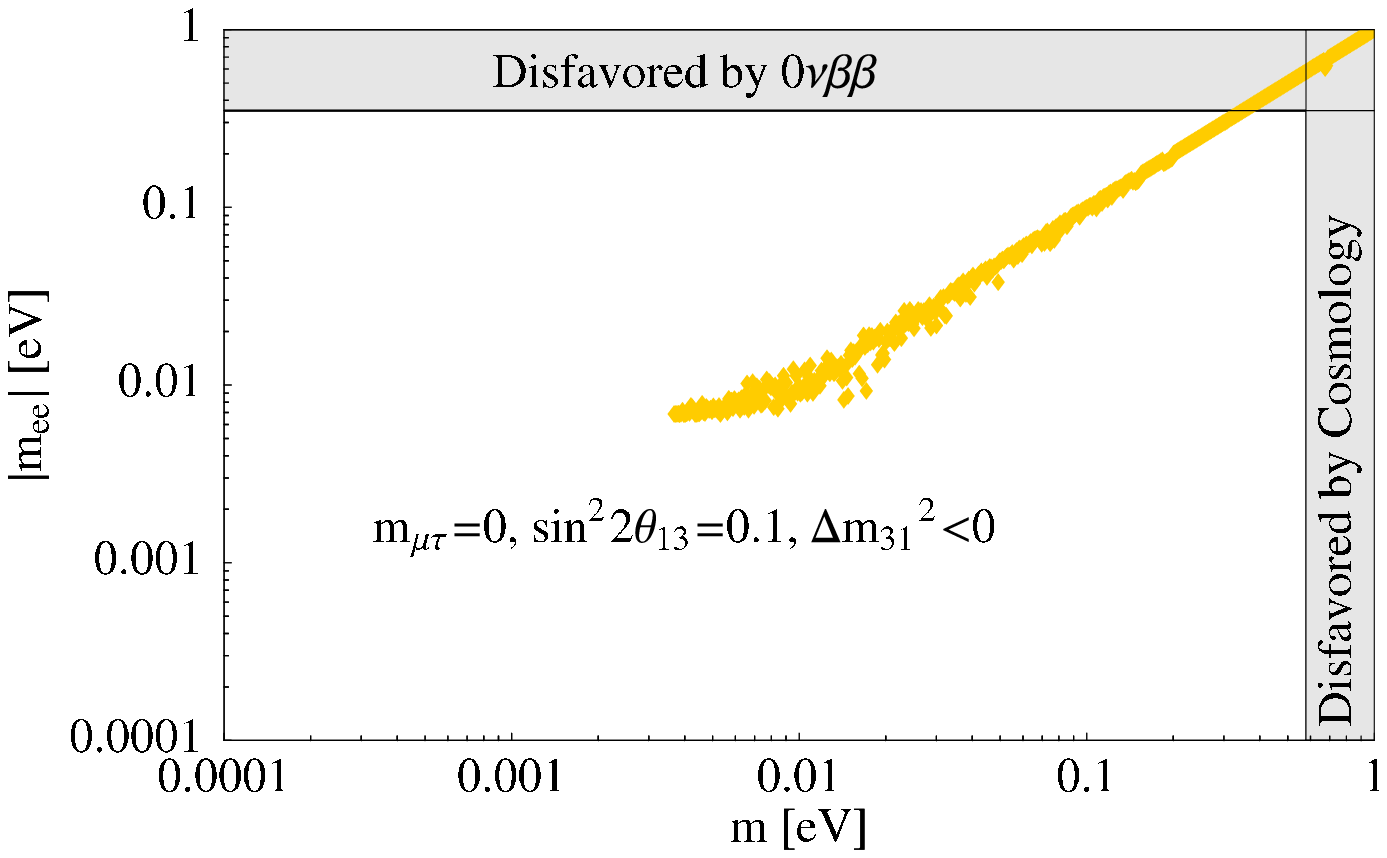,width=8cm,height=6cm}
\caption{\label{fig:mmm2}The effective mass 
$|m_{ee}|$ in case of $m_{\mu\mu}=0$ and $m_{\mu\tau}=0$ for 
$\sin^2 2 \theta_{13}=0.1$. The normal (inverted) mass 
ordering is given on the left (right) side.}
\end{center}
\end{figure}

We see in particular that close-to-maximal atmospheric mixing 
implies small $\sin \alpha$, which indicates 
very little cancellation in the effective mass. 
In Fig.~\ref{fig:mmm2} we give scatter plots of the smallest mass 
against \meff, obtained by demanding $m_{\mu\mu}$ and $m_{\mu\tau}$ 
to vanish and by varying the oscillation parameters within their $3\sigma$ 
ranges. In case of an inverted ordering, $m_{\mu\mu}=0$ is possible 
for all masses, whereas $m_{\mu\tau}=0$ works only for $m_3 \gs 0.01$ eV. 
Note that $m_{ee}=0$ will not be possible if $m_{\mu\mu}$, $m_{\mu\tau}$ 
or $m_{\tau\tau}$ also vanish \cite{2zero,2zero1}. 
For normal ordering, $m_{\mu\mu}=0$ is possible 
when $m_1 \gs 0.02$ eV, 
whereas $m_{\mu\tau}=0$ works only for $m_1 \gs 0.1$ eV. 
The small band of values of \meff~for large masses is a consequence of 
$\sin \alpha \simeq 0$. 

As summarized in Table \ref{tab:FGM}, $m_{\mu\mu}$ and $m_{\mu\tau}$ 
can vanish simultaneously. Also $m_{\mu\mu}$ and $m_{e \mu}$ 
(or $m_{e \tau}$) and $m_{\tau\tau}$ and $m_{e \mu}$ 
(or $m_{e \tau}$) can vanish at the same time. Vanishing of $m_{\mu\tau}$ 
will only be possible if no other entry of $m_\nu$ is zero.

\section{\label{sec:concl}Conclusions and Summary}

We have studied the allowed ranges of the individual mass matrix 
elements in the charged lepton basis. 
Areas of parameter space in which they can vanish have been analyzed, their  
phenomenological consequences have been studied and a limited 
number of correlations has been found. 
Let us summarize the main features:
\begin{itemize}
\item 
the $ee$ element, or the effective mass, can vanish for all values of 
$\theta_{13}$, but only for small values of $m_1$ in the normal 
mass ordering. 
As a function of the smallest neutrino mass, it has from all six 
independent elements the most interesting structure; 
\item 
the $e\mu$ and $e\tau$ entries are very much alike. 
If $\theta_{13}=0$, they cannot vanish. For 
non-zero $\theta_{13}$ there can be a texture zero for all mass values. 
However, the inverted hierarchy and the quasi-degenerate spectrum 
require that $\sin \alpha \simeq 0$, 
which implies little cancellation in the effective mass. 
Consequently, distinguishing the normal from the inverted hierarchy is easier 
in this case;  
\item the $\mu\mu$ and $\tau\tau$ entries are also very much alike. 
In case of $\theta_{13}=0$, vanishing is only possible for a smallest mass 
larger than $\simeq 0.01$ eV. For non-zero $\theta_{13}$ and 
normal ordering, zero entries also require that $m_1 \gs 0.01$ eV, 
whereas for an inverted ordering all mass values allow for complete 
cancellation. For quasi-degenerate masses, $\theta_{23}$ lies below 
$\pi/4$ if $m_{\mu\mu}=0$ and above $\pi/4$ if $m_{\tau\tau}=0$. 
Again, $\sin \alpha \simeq 0$ is required, which translates into 
little cancellation for \meff; 
\item 
the $\mu\tau$ entry cannot vanish if $\theta_{13}=0$ and the masses 
are normally ordered. Non-zero $\theta_{13}$ and normal ordering 
allows for zero $m_{\mu\tau}$ if $m_1 \gs 0.1$ eV, but 
$m_3$ can be larger than a few times $0.001$ eV for an inverted ordering. 
Vanishing in case of NH or IH is not possible. 
Again, for quasi-degenerate neutrinos 
$\sin \alpha \simeq 0$ is required, which translates into 
little cancellation for \meff.
\end{itemize}
We finish by stressing once more that typically the requirement of a 
vanishing element (except for $m_{ee}$, of course) 
leads to little cancellation for the effective mass.

\vspace{0.5cm}
\begin{center}
{\bf Acknowledgments:}\\
\end{center}
This work has been supported by the ``Deutsche Forschungsgemeinschaft'' in the 
``Sonderforschungsbereich 375 f\"ur Astroteilchenphysik'' (A.M.\ and W.R.) 
and under project number RO-2516/3-1 (W.R.). 
W.R.\ would like to thank the ``European Network of Theoretical 
Astroparticle Physics ILIAS/N6'' under contract number RII3-CT-2004-506222
for financial support.

\newpage

\pagestyle{empty}

\begin{table}
\begin{center}
\begin{tabular}{|c|c|}
\hline 
Matrix & Correlation \\
\hline \hline
$\left(\begin{array}{ccc}
    0&0&a\\
    0&b&d\\
    a&d&e
\end{array}\right)$& $ \ba 
\mbox{only NH} \\
|U_{e3}| \simeq \frac{\sqrt{R}}{2} \cot \theta_{23} 
\frac{\sin 2 \theta_{12}}{ \sqrt{\cos 2 \theta_{12}}}  \ea $ \\
\hline
$\left(\begin{array}{ccc}
    0&a&0\\
    a&b&d\\
    0&d&e
\end{array}\right)$&
$ \ba 
\mbox{only NH} \\
|U_{e3}| \simeq \frac{\sqrt{R}}{2} \tan \theta_{23} 
\frac{\sin 2 \theta_{12}}{ \sqrt{\cos 2 \theta_{12}}}  \ea $ \\
\hline \hline
$\left(\begin{array}{ccc}
    a&0&b\\
    0&0&d\\
    b&d&e
\end{array}\right)$& $ \ba \mbox{QD; 
both orderings} \\
|U_{e3}| \simeq \frac{R}{2} \, |\frac{\cot 2\theta_{23} }{\cos \delta }|
\sin 2 \theta_{12} 
\ea $ \\
\hline
 $\left(\begin{array}{ccc}
    a&b&0\\
    b&0&d\\
    0&d&e
\end{array}\right)$& $ \ba \mbox{same as above } 
\ea $ \\
\hline 
$\left(\begin{array}{ccc}
    a&b&0\\
    b&d&e\\
    0&e&0
\end{array}\right)$& $ \ba \mbox{QD; 
both orderings} \\
|U_{e3}| \simeq \frac{R}{2}  \, |\frac{\tan 2\theta_{23} }{\cos \delta }|
\sin 2 \theta_{12}
\ea $ \\
\hline
$ \left(\begin{array}{ccc}
    a&0&b\\
    0&d&e\\
    b&e&0
\end{array}\right)$& $ \ba \mbox{same as above } \ea $ \\
\hline \hline 
$\left(\begin{array}{ccc}
    a&b&d\\
    b&0&e\\
    d&e&0
\end{array}\right)$ & $\ba 
\mbox{QD; 
both orderings} \\ |U_{e3}| \simeq 
\frac{\cot 2 \theta_{12} \, \cot 2 \theta_{23} }{\cos \delta }
\ea $  \\ \hline  
\end{tabular}
\caption{\label{tab:FGM}Mass matrices with two zeros and the resulting 
correlations \cite{2zero,2zero1}.}
\end{center}
\end{table}


\begin{thebibliography}{99} 

\bibitem{APSgen} R.~N.~Mohapatra {\it et al.},
  hep-ph/0412099; hep-ph/0510213. 

\bibitem{PMNS}B. Pontecorvo, 
  Zh. Eksp. Teor. Fiz.\ {\bf 33}, 549 (1957) and {\bf 34}, 247 (1957); 
  Z. Maki, M. Nakagawa and S. Sakata, 
  Prog. Theor. Phys.\ {\bf 28}, 870 (1962).

\bibitem{ich}W.~Rodejohann,
  Phys.\ Rev.\ D {\bf 62}, 013011 (2000); 
J.\ Phys.\ G {\bf 28}, 1477 (2002).  

\bibitem{FS}M.~Frigerio and A.~Y.~Smirnov,
  Nucl.\ Phys.\ B {\bf 640}, 233 (2002); 
Phys.\ Rev.\ D {\bf 67}, 013007 (2003). 

\bibitem{barger}A.~Atre, V.~Barger and T.~Han,
  Phys.\ Rev.\ D {\bf 71}, 113014 (2005). 




\bibitem{kai}K.~Zuber,
  hep-ph/0008080; 
C.~S.~Lim, E.~Takasugi and M.~Yoshimura,
  Prog.\ Theor.\ Phys.\  {\bf 113}, 1367 (2005); 
A.~Ali, A.~V.~Borisov and N.~B.~Zamorin,
  Eur.\ Phys.\ J.\ C {\bf 21}, 123 (2001). 



\bibitem{det}G.~C.~Branco {\it et al.},  
  Phys.\ Lett.\ B {\bf 562}, 265 (2003).  


\bibitem{det1}
 Z.~Z.~Xing,
  Phys.\ Rev.\ D {\bf 69}, 013006 (2004); 
B.~C.~Chauhan, J.~Pulido and M.~Picariello, 
  Phys.\ Rev.\ D {\bf 73}, 053003 (2006). 

\bibitem{trace}
D.~Black, A.~H.~Fariborz, S.~Nasri and J.~Schechter,
  Phys.\ Rev.\ D {\bf 62}, 073015 (2000); 
X.~G.~He and A.~Zee,
  Phys.\ Lett.\ B {\bf 560}, 87 (2003); 
  Phys.\ Rev.\ D {\bf 68}, 037302 (2003); 
  W.~Rodejohann,
  Phys.\ Lett.\ B {\bf 579}, 127 (2004); 
 S.~Nasri, J.~Schechter and S.~Moussa,
  Phys.\ Rev.\ D {\bf 70}, 053005 (2004); 
Phys.\ Rev.\ D {\bf 71}, 093005 (2005). 



\bibitem{FN}C.~D.~Froggatt and H.~B.~Nielsen,
  Nucl.\ Phys.\ B {\bf 147}, 277 (1979).





\bibitem{other}See for instance 
W.~Grimus {\it et al.},  
  Eur.\ Phys.\ J.\ C {\bf 36}, 227 (2004). 






\bibitem{quaaaaark}H.~Fritzsch and Z.~Z.~Xing,
  Prog.\ Part.\ Nucl.\ Phys.\  {\bf 45}, 1 (2000).






\bibitem{2zero}
P.~H.~Frampton, S.~L.~Glashow and D.~Marfatia,
Phys.\ Lett.\ B {\bf 536}, 79 (2002). 

\bibitem{2zero1}
Z.~Z.~Xing,
Phys.\ Lett.\ B {\bf 530}, 159 (2002); 
B.~R.~Desai, D.~P.~Roy and A.~R.~Vaucher,
Mod.\ Phys.\ Lett.\ A {\bf 18}, 1355 (2003);  
W.~L.~Guo and Z.~Z.~Xing,
 Phys.\ Rev.\ D {\bf 67}, 053002 (2003); 
see also S.~K.~Kang and C.~S.~Kim,
  Phys.\ Rev.\ D {\bf 63}, 113010 (2001). 

\bibitem{dirac}
C.~Hagedorn and W.~Rodejohann,
  JHEP {\bf 0507}, 034 (2005). 



\bibitem{hybrid}S.~Kaneko, H.~Sawanaka and M.~Tanimoto,
  JHEP {\bf 0508}, 073 (2005). 

\bibitem{BHP80}S.~M.~Bilenky, J.~Hosek and S.~T.~Petcov,
  Phys.\ Lett.\ B {\bf 94}, 495 (1980); 
  M.~Doi {\it et al.},
  Phys.\ Lett.\ B {\bf 102}, 323 (1981); 
  J.~Schechter and J.~W.~F.~Valle, 
  Phys.\ Rev.\ D {\bf 23}, 1666 (1981). 



\bibitem{valle}
  M.~Maltoni, T.~Schwetz, M.~A.~Tortola and J.~W.~F.~Valle,
  New J.\ Phys.\  {\bf 6}, 122 (2004). 







\bibitem{STP_rev}S.~T.~Petcov,
  New J.\ Phys.\  {\bf 6}, 109 (2004) and references therein. 


\bibitem{HM}H.~V.~Klapdor-Kleingrothaus {\it et al.},
  Eur.\ Phys.\ J.\ A {\bf 12}, 147 (2001); 
C.~E.~Aalseth {\it et al.}  [IGEX Collaboration],
  Phys.\ Rev.\ D {\bf 65}, 092007 (2002).



\bibitem{APSmass}C.~Aalseth {\it et al.},
  hep-ph/0412300 and references therein. 

\bibitem{Tegmark}M.~Tegmark {\it et al.},
  Phys.\ Rev.\ D {\bf 69}, 103501 (2004).


\bibitem{mee0}S.~Pascoli, S.~T.~Petcov and L.~Wolfenstein,
  Phys.\ Lett.\ B {\bf 524}, 319 (2002); 
 Z.~Z.~Xing,
  Phys.\ Rev.\ D {\bf 68}, 053002 (2003);  
S.~Choubey and W.~Rodejohann,
  Phys.\ Rev.\ D {\bf 72}, 033016 (2005). 


\bibitem{LMR}
M.~Lindner, A.~Merle and W.~Rodejohann,
  Phys.\ Rev.\ D {\bf 73}, 053005 (2006). 

\bibitem{steen}For an overview see S.~Hannestad,
  Nucl.\ Phys.\ Proc.\ Suppl.\  {\bf 145}, 313 (2005), hep-ph/0412181.


\bibitem{mutau}C.~S.~Lam,
  Phys.\ Lett.\ B {\bf 507}, 214 (2001);
 W.~Grimus and L.~Lavoura,
  JHEP {\bf 0107}, 045 (2001);
 E.~Ma,
    Phys.\ Rev.\ D {\bf 66}, 117301 (2002);
  P.~F.~Harrison and W.~G.~Scott,
  Phys.\ Lett.\ B {\bf 547}, 219 (2002);
an incomplete list of more recent studies is:
  Y.~Koide,
  Phys.\ Rev.\ D {\bf 69}, 093001 (2004);
W.~Grimus {\it et al.},
  Nucl.\ Phys.\ B {\bf 713}, 151 (2005);
 R.~N.~Mohapatra,
  JHEP {\bf 0410}, 027 (2004);
 I.~Aizawa, T.~Kitabayashi and M.~Yasue,
  Phys.\ Rev.\ D {\bf 72}, 055014 (2005);
C.~S.~Lam,
  Phys.\ Rev.\ D {\bf 71}, 093001 (2005);
R.~N.~Mohapatra and W.~Rodejohann,
  Phys.\ Rev.\ D {\bf 72}, 053001 (2005);
A.~S.~Joshipura,
  hep-ph/0512252; 
Y.~H.~Ahn {\it et al.}, 
  hep-ph/0602160.

\end{thebibliography}
\end{document}